\documentclass[%
 reprint,
superscriptaddress,
 amsmath,amssymb,
 aps,
]{revtex4-2}

\usepackage{graphicx}
\usepackage{dcolumn}
\usepackage{multirow}
\usepackage{bm}
\usepackage{cleveref}

\begin{document}

\preprint{APS/123-QED}

\title{Mapping recrystallization trajectories in GaAs using\\ latent space diffraction analysis}

\author{Ellis Rae Kennedy}
\email{erk89@cornell.edu}%
\affiliation{Department of Chemistry and Chemical Biology, Cornell University, Ithaca, NY, USA}%

\author{Kwanghwi Je}
\affiliation{Department of Chemistry and Chemical Biology, Cornell University, Ithaca, NY, USA}%

\author{Erik Thiede}
\affiliation{Department of Chemistry and Chemical Biology, Cornell University, Ithaca, NY, USA}%

\date{\today}

\begin{abstract}
Recrystallization in disordered solids proceeds through a sequence of local structural rearrangements that are difficult to resolve using conventional diffraction analysis. In amorphous and partially ordered materials, subtle variations in diffuse scattering, short-range order, and defect-mediated symmetry emergence encode the pathways through which ordering initiates and propagates. Here, we introduce a latent space framework for mapping these pathways directly from \textit{in situ} 4D-STEM diffraction data. A convolutional autoencoder provides a compact representation of structural motifs, and unsupervised clustering identifies recurring microstructural states, including amorphous, paracrystalline, crystalline, twinned, and hybrid intermediates. By tracking these states across temperature, we construct phase trajectory models that reveal the topology of the recrystallization landscape, including metastable basins, branching pathways, hybrid states, and temperature-dependent reorganizations of accessible states.

Applied to ion irradiated GaAs, this approach uncovers two distinct recrystallization regimes separated by a transition near 250\textdegree{}C. At low temperature, recrystallization is growth-dominated and dominated by the persistence of amorphous and crystalline states. At high temperature, the transformation landscape reorganizes: hybrid and faulted states become metastable precursors to twinning, polycrystalline regions stabilize, and twinned structures emerge as dominant end states. The latent space representation also identifies amorphous patterns with weak symmetry signatures that precede recrystallization. This reveals structural precursors to ordering that are not captured by conventional descriptors give new insights into how recrystallization is initiated. These results demonstrate that integrating \textit{in situ} 4D-STEM with nonlinear dimensionality reduction and statistical trajectory analysis can uncover phase transformation pathways in complex materials and  reveal how nanoscale heterogeneity, metastability, and defect-mediated intermediates shape the evolution of structure under thermal driving.
\end{abstract}

\maketitle

\section{Introduction}

Recrystallization is a primary mechanism through which disordered solids recover structure. Here, recrystallization is defined as the transformation of an amorphous solid into a more ordered crystalline state, resulting in a recovered structure~\cite{zhao2021tellurium,velisa2025revealing}. Despite being observed across material systems, the microscopic processes that govern this evolution remain difficult to quantify. In many materials, including semiconductors, this amorphous-to-crystalline transformation is neither spatially uniform nor governed by a single pathway~\cite{treacy2012local,rosset2025signatures}. Instead, it proceeds through a sequence of local events, including interface motion, defect rearrangement, symmetry emergence, and the formation or dissolution of short-range order, that are sensitive to the initial disorder landscape and driving force~\cite{wu2024elemental,adam2021analyzing,rudawski2015semiconductors}. Amorphous phases often retain medium-range ordering structural motifs that influence how ordering initiates, but these signatures are subtle and frequently obscured in conventional characterization techniques and diffraction analysis~\cite{voyles2001structure,sorensen2020revealing,kennedy2024exploring}.

Four-dimensional scanning transmission electron microscopy (4D-STEM) provides a route to resolving this complexity. In 4D‑STEM, a rastered electron probe produces a diffraction pattern at every pixel, enabling quantitative mapping of structural evolution with nanometer precision~\cite{gammer2018local,ophus2019four,voyles2002fluctuation}. Each diffraction pattern encodes local symmetry, orientation, and disorder, offering a nanoscale record of structural evolution during \textit{in situ} heating~\cite{pekin2019direct,savitzky2021py4dstem}. The resulting datasets are high-dimensional, noisy, and contain overlapping crystalline and diffuse features. However, these datasets are often analyzed using scalar descriptors, such as peak intensities or radial averages, which fail to capture the continuous, multi-pathway nature of recrystallization, particularly when intermediate states differ only by subtle changes in symmetry or diffuse scattering.

To address these challenges, we develop a latent space framework for mapping recrystallization pathways directly from 4D-STEM data. A convolutional autoencoder (CAE) learns a compact representation of diffraction patterns that preserves physically meaningful variations in disorder, symmetry, and orientation. Unsupervised clustering of the latent vectors identifies recurring microstructural states, including amorphous, paracrystalline, crystalline, twinned, polycrystalline, and hybrid intermediates. By tracking these states across temperature, we construct transition matrices that reveal transition pathways, metastable basins, and temperature-dependent reorganization.

\begin{figure}[t]
\centering
\includegraphics[width = 0.44\textwidth]{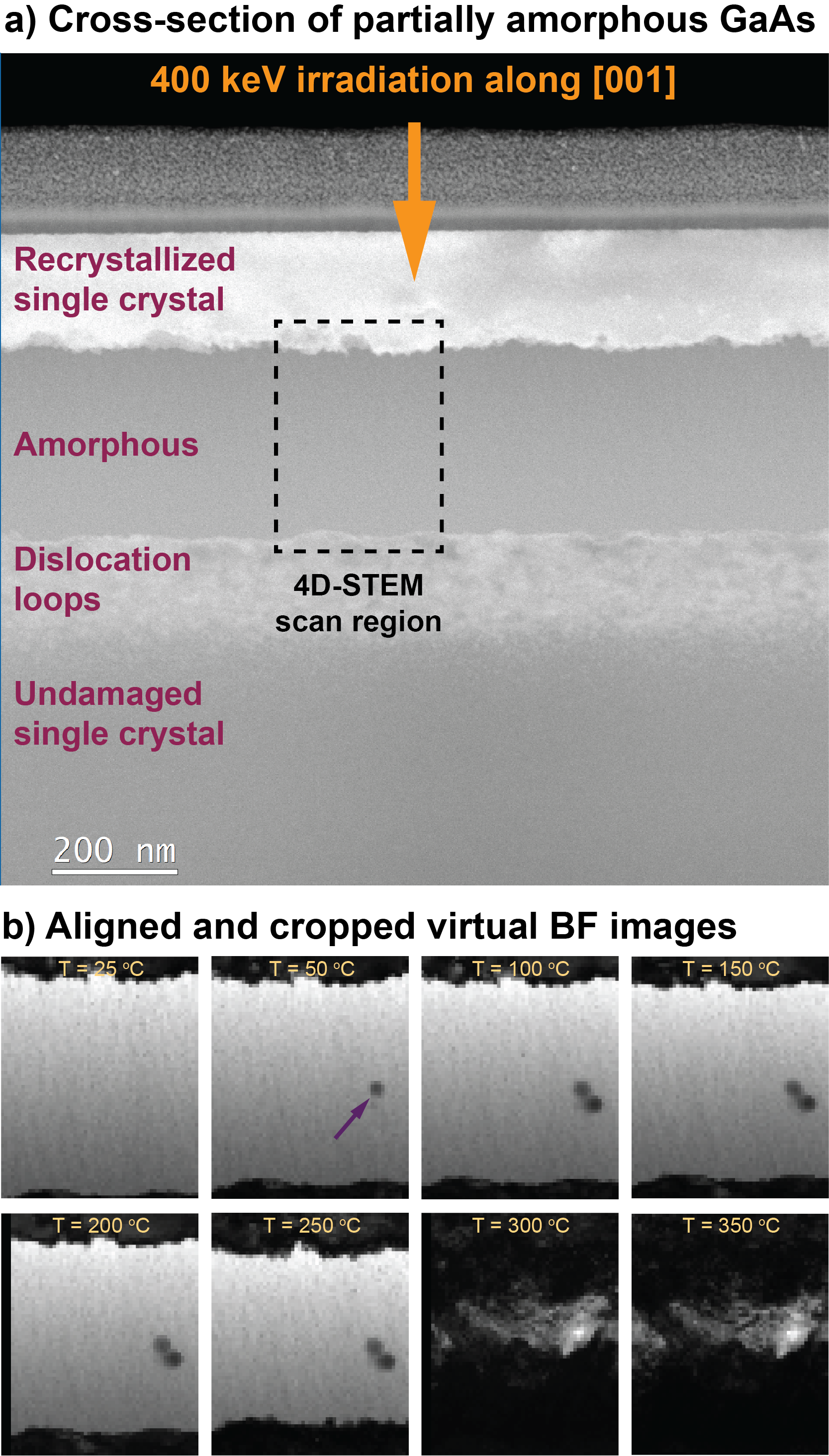}
  \caption{\textbf{Sample geometry and temperature-resolved 4D-STEM imaging.} (a) Cross-sectional TEM micrograph of partially amorphous GaAs following 400 keV Ne$^{++}$ irradiation along the [001] direction. Distinct regions of recrystallized single crystal, amorphous GaAs, dislocation loops, and undamaged substrate are visible. The dashed box indicates the region used for temperature series 4D-STEM acquisition. (b) Aligned and cropped virtual bright-field (BF) images from the same scan region at temperatures between 25\textdegree{}C and 350\textdegree{}C. Contrast changes reflect the progressive loss of the amorphous layer and the emergence of crystalline and defected microstructures during \textit{in situ} annealing.}

  \label{fig:overview}
\end{figure}

We apply this approach to ion irradiated GaAs (Figure~\ref{fig:overview}), a technologically important III-V semiconductor whose amorphous phase retains measurable structural memory and whose recrystallization exhibits distinct low- and high-temperature regimes~\cite{allen1960gallium}. GaAs is a quintessential material for photovoltaics and optoelectronic devices with various applications in non-equilibrium environments, such as in the temperature and irradiation extremes of outer space~\cite{liu2025review,srour1988radiation,pearton2021review}. Here, the amorphous phase is produced by disordering of an initially crystalline lattice using Ne$^{++}$ ions. GaAs provides a well-defined platform for studying structural recovery: its amorphous matrix contains short-range order inherited from the crystalline parent; its nucleation barrier produces a clear transition between growth-dominated and nucleation-dominated regimes; and its high-temperature behavior is dominated by the formation of \{111\} nanotwin networks similar to those observed in other cubic semiconductors~\cite{velisa2025revealing,gibson1997diminished,gerthsen1993stacking}.

\begin{figure*}[t]
\centering
\includegraphics[width = 0.97\textwidth]{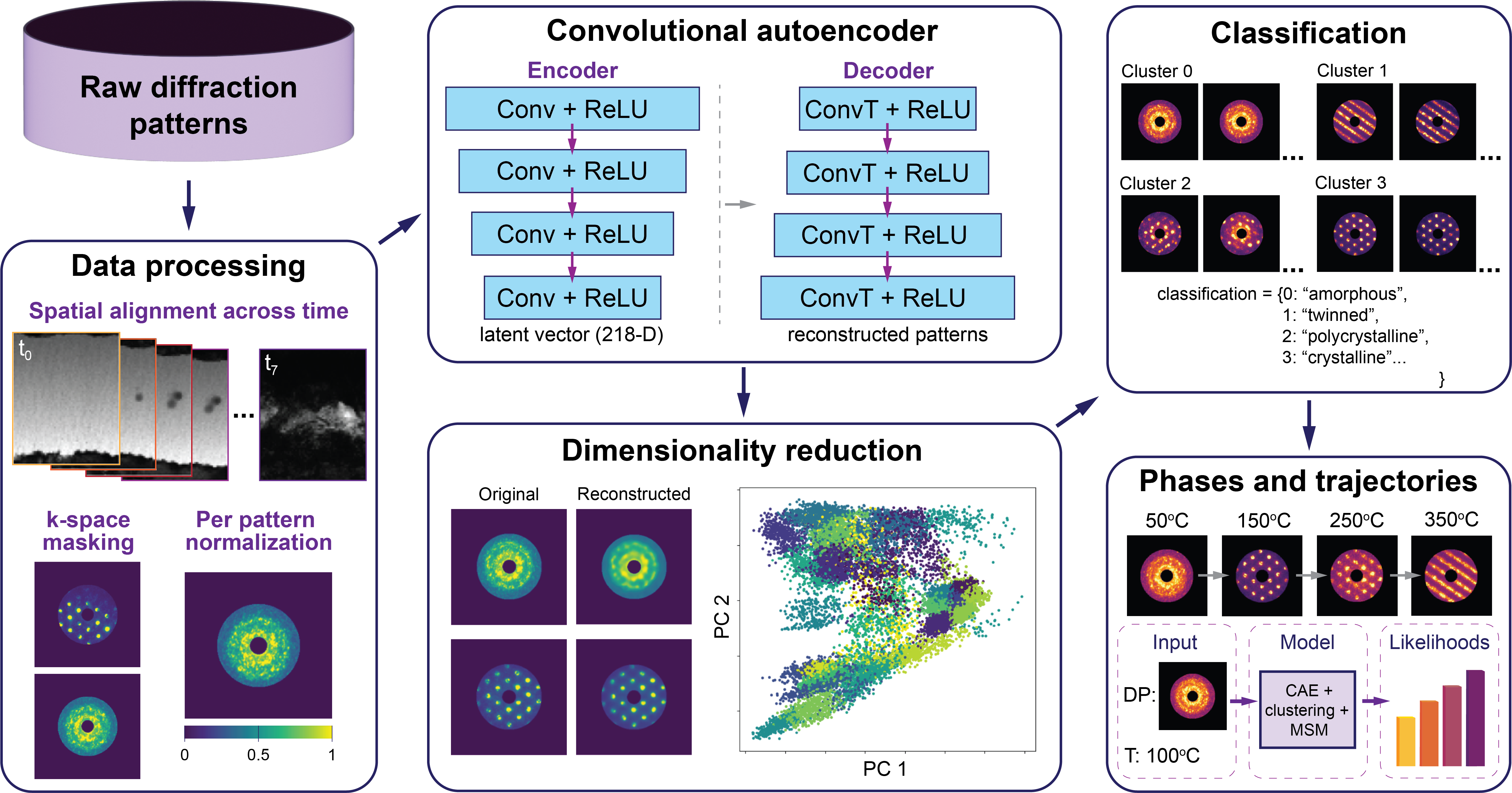}
  \caption{Workflow for extracting structural phases and transition pathways from \textit{in situ} diffraction data. Raw diffraction patterns undergo preprocessing, including spatial alignment across time, reciprocal space masking to isolate physically meaningful scattering, and per-pattern normalization. The cleaned patterns are encoded using a convolutional autoencoder, which compresses each pattern into a 218-dimensional latent vector and reconstructs it through a symmetric decoder. Latent representations are projected into a low-dimensional space for visualization and clustered to identify distinct structural states. These cluster assignments form the basis for measuring transition probabilities and reconstructing temperature-dependent phase trajectories. The combined CAE, clustering, and transition matrix pipeline produces phase labels and probabilistic pathways describing the evolution of local structure during heating.}
  \label{fig:workflow}
\end{figure*}

Using the latent space framework shown in Figure~\ref{fig:workflow}, we uncover previously unresolved features of the nanoscale recrystallization mechanism in GaAs. At low temperature, both amorphous and crystalline regions remain kinetically trapped. Although crystalline order is energetically favored, the amorphous network lacks the mobility required for interface motion or defect rearrangement, preventing the system from accessing lower energy ordered configurations~\cite{kennedy2026critical}. At high temperature, the transformation landscape reorganizes: hybrid and faulted states become metastable basins and precursors to twinning, polycrystalline regions stabilize, and twinned structures emerge as dominant end states. The latent space also reveals amorphous patterns with weak symmetry signatures that precede recrystallization, indicating structural precursors to ordering that are not captured by conventional descriptors. Most importantly, we are able to track these changes on the nanometer length scale to precisely understand the temporal-spatial kinetics of the system.

These results establish a physics-informed, data-driven framework for quantifying phase transformation pathways. Coupling latent space structural classification with trajectory analysis enables direct mapping of system dynamics and reveals how nanoscale heterogeneity, metastable states, and defect‑mediated intermediates steer structural evolution during thermal activation.

\section*{Results}
 
\begin{figure*}[t]
\centering
\includegraphics[width=0.95\textwidth]{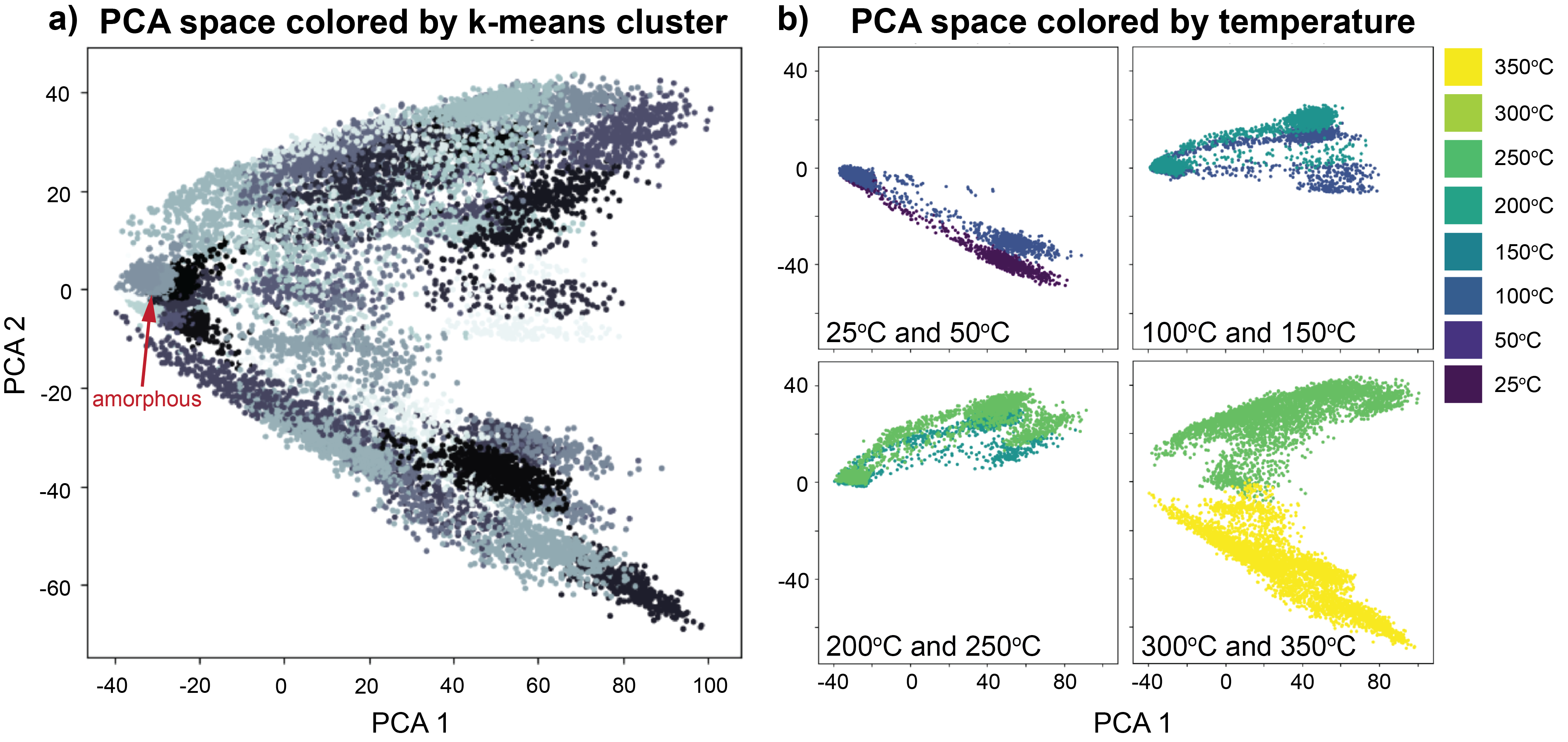}
\caption{\textbf{PCA embedding of all diffraction pattern latent vectors.} (a) Points colored by k-means cluster assignment in latent space. There are 50 clusters. (b) The same PCA space colored by temperature, showing a progression from low- and mid-temperature amorphous patterns to high-temperature ordered states, with the amorphous basin occupying a distinct low PCA 1 region. The population shifts from amorphous dominated at low temperature to ordered states at high temperature; amorphous patterns become negligible by the highest temperature steps.}
\label{fig:pcaspace}
\end{figure*}

We train a convolutional autoencoder (CAE) with pre-processed \textit{in situ} datasets from GaAs heated in a TEM to obtain latent space representations of the diffraction patterns (Figure~\ref{fig:workflow}). Clustering performed directly on raw or normalized patterns produces poorly separated groups and inconsistent boundaries (see Kennedy \textit{et al.}\cite{kennedy2026critical}), whereas the CAE yields a latent representation in which diffraction patterns organize according to physically meaningful variations in disorder, symmetry, and orientation (Figure~\ref{fig:pcaspace})~\cite{ivanov2024autoencoder,chen2017outlier}. Principal component analysis (PCA) of the 218-dimensional latent vectors followed by $k$-means clustering reveals a structured manifold with distinct regions corresponding to amorphous (A), crystalline (C), recrystallized (R), twinned (T), polycrystalline (P), and hybrid (H) states~\cite{yoo2024unsupervised}. The phase classification of each cluster was manually assigned (see Table I in the Supplemental Information). The recrystallized phase is structurally identical to the crystalline phase but differs by less than $1^\circ$ in orientation due to epitaxial regrowth\cite{kennedy2026critical}; because this distinction is negligible for the purposes of trajectory analysis, we treat C and R together as a single ordered basin, C$^\ast$, in later sections. In Figure~\ref{fig:pcaspace}b, the patterns are plotted in PCA space and colored by temperature, highlighting how low-temperature amorphous features occupy distinct tightly bound PCA coordinates compared to the high-temperature ordered features. This latent space representation constitutes a fundamentally new way of characterizing \textit{in situ} diffraction data: instead of relying on manually defined metrics or peak-based descriptors, the CAE learns a structural manifold that captures disorder, symmetry, and orientation with nanometer spatial resolution.

To identify microstructural states within the latent manifold, $k$-means clustering is applied to the latent vectors and then structure assignments are made for all patterns across temperature steps to define robust macrostates. Here, a macrostate refers to a recurrent, statistically coherent microstructural configuration in latent space that groups together patterns sharing similar disorder, symmetry, and orientation. Manual labeling of the unsupervised clusters is used to assign a structural classification to each cluster. This includes hybrid (H) states that occupy boundary regions between major phases, forming continuous bridges between amorphous and crystalline (A/C), crystalline and twinned (C/T), crystalline and polycrystalline (C/P), and twinned and polycrystalline (T/P) configurations. These hybrid states are not artifacts of clustering; rather, they consist of well-defined clusters of patterns that represent hybrids between clearly defined phases. They are initially considered as unique combinations of two structures, but later combined into a general H state. In this system, we distinguish between local stacking‑faulted configurations (“faulted”), which represent short Wurtzite-like interruptions in the \{111\} stacking sequence, and extended twinned domains (“twinned”), which form when these faults propagate to create a coherent twin boundary. The C/T hybrids correspond to the former and serve as intermediates that funnel the system toward the latter.

Representative reconstructions (Figure S1 of the Supplemental Information) confirm that CAE-encoded patterns match the corresponding experimental patterns with high fidelity once the model has trained for approximately 50 epochs. Mapping phase labels back onto the real space scan grid reveals the spatial progression of recrystallization: at low temperature, the crystalline-amorphous interface advances slowly with isolated pockets of partial ordering forming ahead of the front, whereas at high temperature the amorphous matrix collapses rapidly and twinned domains form and propagate throughout the field of view. Hybrid states appear predominantly at moving interfaces and in regions undergoing structural rearrangement, consistent with their role as intermediates between major phases~\cite{treacy2012local,porter2009phase}. A hybrid state therefore corresponds to a local boundary between two microstructures, reflecting the presence of an interface where elements of both macrostates coexist. This work provides the first nanometer-resolved, \textit{in situ} observation of high-temperature recrystallization pathways in an amorphous solids, revealing structural intermediates that have not been accessible in \textit{ex situ} or ensemble averaged measurements.

\begin{figure}[b]
\centering
\includegraphics[width=0.49\textwidth]{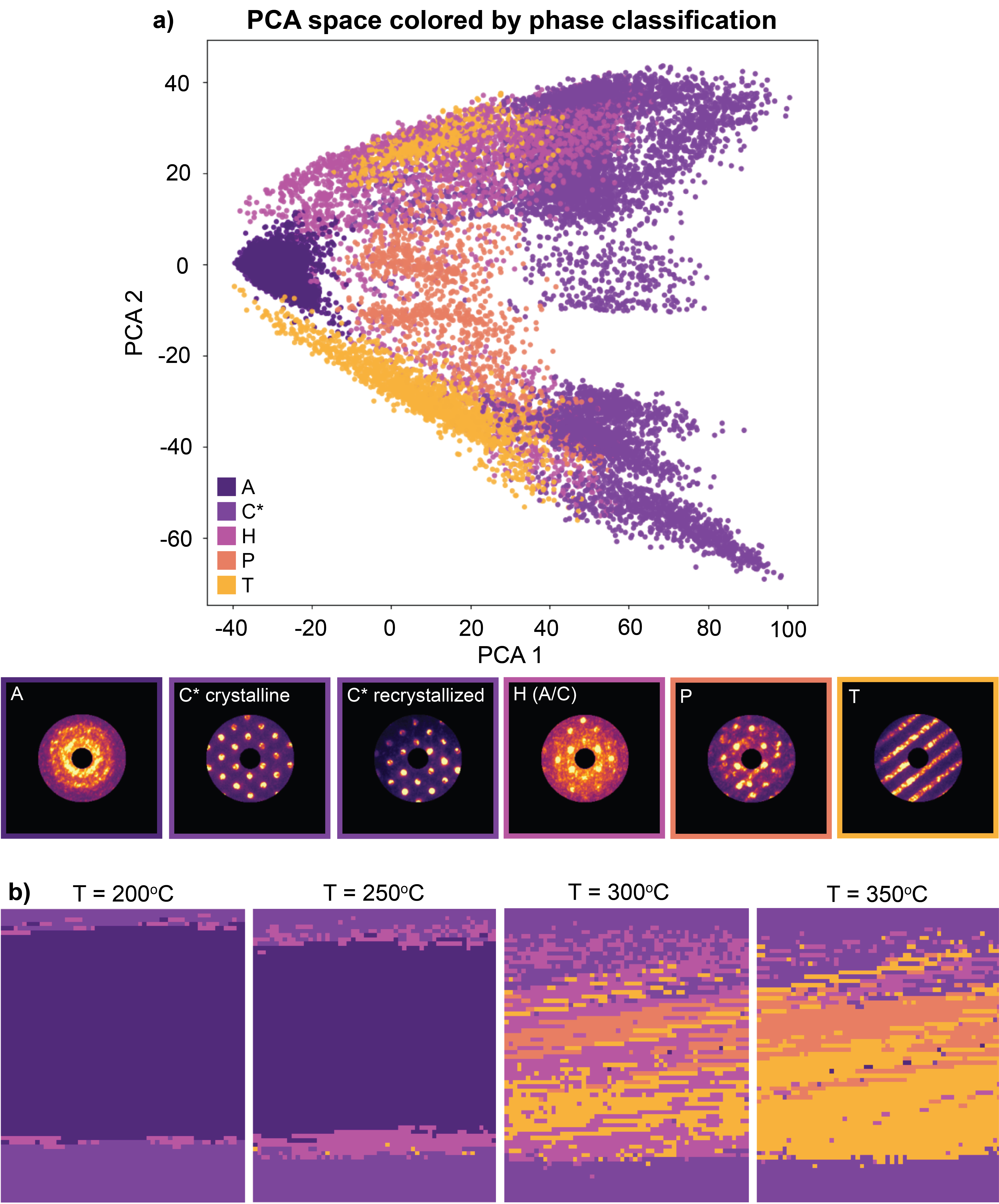}
\caption{\textbf{PCA space colored by classification and phase maps highlighting structural reorganization near 250\textdegree{}C.} (a) Two-dimensional PCA plot of latent vectors, colored by phase classification (A, C$^\ast$, H, P, T). Distinct clusters reflect how the autoencoder derived latent space organizes diffraction patterns according to their underlying structural phase. Representative diffraction patterns are shown for each phase. (b) These temperature steps include the transition from a nucleation- to growth-limited regime, where the system undergoes its most rapid structural reorganization. The maps show the disappearance of low-temperature amorphous and C$^\ast$ regions and the first appearance of hybrid (H), polycrystalline (P), and twinned (T) structures.}
\label{fig:phasemap}
\end{figure}

Having established the structural organization of the latent space, we next investigate how these phases evolve with temperature. The evolution of phase populations across temperature is shown in Table~\ref{tab:counts} and evinces that the amorphous (A) and crystalline (C$^\ast$) phases dominate the microstructure throughout the low-temperature regime, with only small fractions of hybrid (H), polycrystalline (P), or recrystallized (R) patterns present. Above 250\textdegree{}C, however, the amorphous population collapses and hybrid, polycrystalline, and twinned states emerge sharply. This is also evident in Figure~\ref{fig:phasemap}, where the phase populations change dramatically above 250\textdegree{}C. Line plots of the phase counts as a function of temperature is shown in Figure S2 of the Supplemental Information. To quantify how these populations evolve and develop a mechanistic understanding of the system, we construct a temperature-indexed transition matrix (Tables 2 and 3 of the Supplemental Information) from single-pattern resolved trajectories across consecutive temperature steps, enabling us to examine how local nanometer-scale structural motifs reorganize during heating. The combination of trajectories with a learned structural manifold, enables a reconstruction of temperature-dependent recrystallization pathways with a level of mechanistic detail that has not previously been achievable with conventional \textit{in situ} electron microscopy workflows.

\begin{table*}[t]
\centering
\renewcommand{\arraystretch}{1.25}
\setlength{\tabcolsep}{6pt}
\begin{tabular}{c|cccccc}
\hline
Temp & Amorphous (A) & Crystalline (C) & Hybrid (H) & Polycrystalline (P) & Recrystallized (R) & Twinned (T) \\
\hline
25\textdegree{}C & 3263 (0.82) & 0 (0.00) & 80 (0.02) & 0 (0.00) & 647 (0.16) & 0 (0.00) \\
50\textdegree{}C & 3269 (0.82) & 18 (0.01) & 104 (0.03) & 0 (0.00) & 599 (0.15) & 0 (0.00) \\
100\textdegree{}C & 2671 (0.75) & 915 (0.23) & 104 (0.03) & 0 (0.00) & 0 (0.00) & 0 (0.00) \\
150\textdegree{}C & 2886 (0.72) & 1002 (0.25) & 102 (0.03) & 0 (0.00) & 0 (0.00) & 0 (0.00) \\
200\textdegree{}C & 2810 (0.70) & 1071 (0.27) & 109 (0.03) & 0 (0.00) & 0 (0.00) & 0 (0.00) \\
250\textdegree{}C & 2529 (0.63) & 1051 (0.26) & 403 (0.10) & 0 (0.00) & 0 (0.00) & 7 (0.00) \\
300\textdegree{}C & 3 (0.00) & 1367 (0.34) & 1458 (0.37) & 416 (0.10) & 0 (0.00) & 746 (0.19) \\
350\textdegree{}C & 7 (0.00) & 0 (0.00) & 242 (0.06) & 727 (0.18) & 1498 (0.38) & 1516 (0.38) \\
\hline
\end{tabular}
\caption{Counts and fractions of all detected phases at each temperature.}
\label{tab:counts}
\end{table*}

We observe a substantial reorganization of transition probabilities between the low-temperature (25--200\textdegree{}C) and high-temperature (250--350\textdegree{}C) regimes (Table~\ref{tab:msm}). At low temperature, the amorphous phase remains highly stable (A$\rightarrow$A = 0.91), with only small fractions transitioning into C$^\ast$ (A$\rightarrow$C$^\ast$ = 0.08) or hybrid states (A$\rightarrow$H = 0.02). The ordered basin exhibits similarly strong self-retention (C$^\ast\rightarrow$C$^\ast$ = 0.79), indicating suppressed defect mobility and limited lattice reorganization. Hybrid states behave as short-lived intermediates, predominantly dissolving back into A (H$\rightarrow$A = 0.51) or annealing into C$^\ast$ (H$\rightarrow$C$^\ast$ = 0.43). Polycrystalline and twinned states are absent in this regime, consistent with a growth-dominated landscape in which A and C$^\ast$ act as deep basins and all other states behave as transient intermediates.

In contrast, above 250\textdegree{}C the phase-count table (Table~\ref{tab:counts}) shows a collapse of the amorphous population and the rapid emergence of hybrid (H), polycrystalline (P), and twinned (T) states. This reorganization is also evident in the virtual bright-field images (Figure~\ref{fig:overview}). To quantify how transitions occur between phases in this regime, we compute the full transition matrix for 250--350\textdegree{}C (Table~\ref{tab:msm}), using the combined ordered basin C$^\ast$ for crystalline and recrystallized states.

Above 250\textdegree{}C, the amorphous phase becomes completely unstable (A$\rightarrow$A = 0.00), transitioning primarily into hybrid (A$\rightarrow$H = 0.45), twinned (A$\rightarrow$T = 0.27), polycrystalline (A$\rightarrow$P = 0.17), and ordered (A$\rightarrow$C$^\ast$ = 0.12) states. These pathways reflect the onset of nucleation-dominated recrystallization, in which the amorphous matrix rapidly reorganizes into ordered configurations.

\begin{table*}[!]
\centering
\renewcommand{\arraystretch}{1.2}
\setlength{\tabcolsep}{9pt}
\label{tab:msm_cstar}

\begin{tabular}{c|ccccc|c|ccccc}
\hline
 & \multicolumn{5}{c|}{\textbf{25--200\textdegree{}C}} 
 &  & \multicolumn{5}{c}{\textbf{250--350\textdegree{}C}} \\
\textbf{From/To} 
& A & C* & H & P & T
&  & A & C* & H & P & T \\
\hline

A 
& 0.91 & 0.08 & 0.02 & 0.00 & 0.00
&  & -- & 0.12 & 0.45 & 0.17 & 0.27 \\

C* 
& 0.16 & 0.79 & 0.06 & 0.00 & 0.00
&  & 0.00 & 0.94 & 0.04 & 0.01 & 0.02 \\

H 
& 0.51 & 0.43 & 0.06 & 0.00 & 0.00
&  & 0.00 & 0.15 & 0.23 & 0.15 & 0.47 \\

P 
& -- & -- & -- & -- & --
&  & 0.00 & 0.01 & 0.00 & 0.94 & 0.06 \\

T 
& -- & -- & -- & -- & --
&  & 0.00 & 0.01 & 0.08 & 0.05 & 0.86 \\

\hline
\end{tabular}
\caption{\textbf{Collapsed phase transition probabilities for low- and high-temperature regimes.}
Crystalline (C) and recrystallized (R) states are combined into a single basin C\*. 
Left: 25--200\textdegree{}C (Temp Steps 0--4). Right: 250--350\textdegree{}C (Temp Steps 5--7). 
At low temperature, A and C\* are strongly self-stabilizing, while H exhibits mixed outgoing pathways. 
At high temperature, C\* becomes highly absorbing, P forms a deep metastable basin, and T dominates the terminal configuration space.}
\label{tab:msm}
\end{table*}

The ordered basin remains strongly self-stabilizing (C$^\ast\rightarrow$C$^\ast$ = 0.94) but exhibits finite transitions into hybrid (C$^\ast\rightarrow$H = 0.04), polycrystalline (C$^\ast\rightarrow$P = 0.01), and twinned (C$^\ast\rightarrow$T = 0.02) states, indicating that defect mobility and stacking fault formation become thermally accessible. Hybrid states, indicative of interfaces, become major conduits for high-temperature reorganization, transitioning primarily into T (H$\rightarrow$T = 0.47), C$^\ast$ (H$\rightarrow$C$^\ast$ = 0.15), and P (H$\rightarrow$P = 0.15), consistent with faulted or partially ordered structures feeding into grain growth and twin formation.

Polycrystalline regions form a deep metastable basin (P$\rightarrow$P = 0.94), with fewer transition pathways into T (P$\rightarrow$T = 0.06), reflecting grain coalescence and boundary migration at high temperature. Twinned regions are the most stable high-temperature configuration (T$\rightarrow$T = 0.86), with only minor transitions into hybrid (T$\rightarrow$H = 0.08), polycrystalline (T$\rightarrow$P = 0.05), or ordered (T$\rightarrow$C$^\ast$ = 0.01) states.

These transitions show that above 250\textdegree{}C the recrystallization landscape reorganizes into a fault-driven, multi-pathway regime in which hybrid, polycrystalline, and ordered states all feed into twinning. Twinned structures thus emerge as the dominant terminal state, acting as the primary sink for transitions out of A, C$^\ast$, H, and P.

\begin{figure*}[t]
\centering
\includegraphics[width=0.85\textwidth]{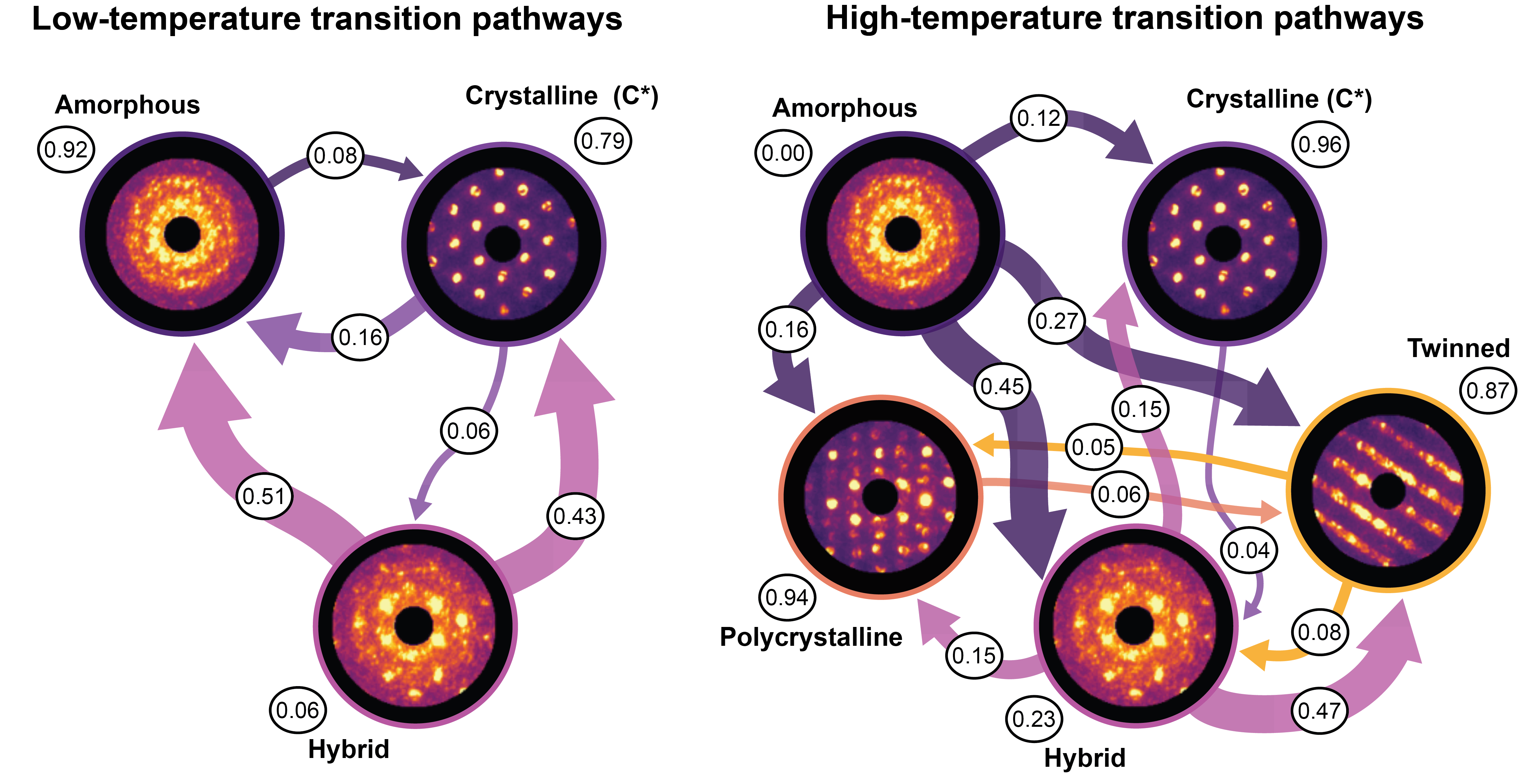}
\caption{\textbf{Temperature-dependent phase transition pathways derived from the phase transition matrix.}
Directed edges indicate transitions with probability greater than 3\% and supported by at least 10 observed pixel trajectories between consecutive temperature steps. Arrow thickness is proportional to transition probability. Left: Low-temperature regime (25--200\textdegree{}C), dominated by persistent A and C phases with weak inter-phase pathways. Right: High-temperature regime (250--350\textdegree{}C), where A becomes unstable, hybrid states form a metastable basin, and twinned and crystalline phases dominate the terminal states.}
\label{fig:transitions}
\end{figure*}

To visualize these changes, we construct a transition network (Figure~\ref{fig:transitions}), applying a 3\% probability cutoff and requiring at least 10 observed transitions. This thresholding removes anomalous or sparsely sampled transitions that likely arise from classification noise rather than genuine structural kinetics. Arrow thickness encodes transition probability, and node values reflect phase stability through their self-transition probabilities. The resulting diagrams highlight the emergence of new high-temperature pathways and the suppression of low-temperature ones, revealing a clear bifurcation between nucleation-dominated and growth-dominated recrystallization dynamics~\cite{doherty1985nucleation}. In irradiated amorphous GaAs, the low-temperature regime is effectively nucleation‑limited because the low-mobility amorphous network suppresses the atomic rearrangements required to form critical nuclei, whereas at high temperature mobility increases and the transformation becomes limited by interface motion and defect-mediated growth rather than nucleus formation~\cite{shao2014effect,janish2020insitu}. Because transitions from multiple temperature steps are aggregated within each regime, these networks should be interpreted as Markov-like models that capture dominant pathways without assuming strict memorylessness at each individual temperature~\cite{metzner2008transition}.

Hybrid states occupy boundary regions and adjacent lobes in latent space and encode mixtures of structural motifs that arise during local structural rearrangement. To resolve their mechanistic roles, we decomposed the hybrid phase (H) into subclusters using $k$-means applied to all hybrid state latent vectors. The resulting clusters span a range of physically interpretable intermediates: two amorphous-crystalline (A/C) hybrids (clusters~40 and~47), one crystalline-polycrystalline (C/P) hybrid (cluster 37), four crystalline-twinned (C/T) hybrids (clusters~25, 34, 43, and~44), one twinned-polycrystalline (T/P) hybrid (cluster 9), and one recrystallized-twinned (R/T) hybrid (cluster 14). Their temperature distributions reveal a clear temperature-dependent separation: A/C hybrids span the full temperature range, with cluster 47 concentrated at the earliest temperatures (mean~0.53), whereas the C/T, C/P, T/P, and R/T hybrids emerge only at high temperature (250--350\textdegree{}C), where defect mobility and fault reorganization are maximal. A complete list of the cluster labels is provided in Table I of the Supplemental Information. 

Tracking each hybrid labeled pixel to the subsequent temperature step reveals distinct mechanistic roles for these intermediates. Transition destination probabilities for the hybrid clusters are shown in Figure~\ref{fig:hybrid}a and Figure S3 of the Supplemental Information. Early A/C hybrids (cluster 47) predominantly dissolve into the amorphous phase (A = 0.66), with a minority transitioning into recrystallized order (R = 0.29), consistent with subcritical nuclei. In contrast, the mid-temperature A/C hybrid (cluster 40) exhibits a strong bias toward crystallization (C = 0.56), with additional flux toward fault formation (C/T = 0.16) and dissolution (A = 0.26). 

The C/T hybrids form the dominant faulted intermediates: cluster 34 transitions to T with probability 0.46, cluster 43 with 0.57, and cluster 44 with 0.49, while cluster 25 feeds primarily into R/T (0.51) and R (0.27). These pathways identify C/T hybrids as the principal precursors to twinning, with increasing determinism at higher temperature. The C/P hybrid (cluster 37) transitions overwhelmingly to P (0.81), reflecting grain-boundary-rich regions that coarsen into polycrystalline grains. The T/P hybrid (cluster 9) transitions primarily to T (0.83), indicating that high-defect polycrystalline regions either sharpen into twins or coarsen into P. This behavior is consistent with the thermodynamic driving force for grain growth, in which reducing interfacial free energy favors the elimination of misoriented boundaries and promotes the development of lower‑energy twin configurations. The transition network helps constrain the origin of annealing twins. Twinned domains rarely appear directly from the amorphous phase; instead, they arise predominantly through faulted C/T intermediates and, to a lesser extent, from polycrystalline configurations. This sequence indicates that twinning is a late-stage, growth-mediated process that proceeds through the propagation of stacking faults and subsequent grain-boundary reorganization~\cite{lin2015observation}. Finally, the R/T hybrid (cluster 14) appears only at the final temperature step and exhibits no outgoing transitions, acting as a terminal high-temperature state.

The geometric structure of these pathways is evident in PCA projections of the latent space (Figure~\ref{fig:hybrid}b). A/C hybrids form lobes that flow toward A, C, or T, revealing that subtle latent space directions encode whether a partially ordered region will dissolve, crystallize, or fault. The C/T hybrids occupy a narrow ridge that flows almost exclusively to T, consistent with a faulted-to-twinned pathway. This coterie of high-temperature hybrids (C/P, T/P, and R/T) clusters tightly and exhibit no low-temperature precursors, reflecting their terminal character.

To quantify the fate of A/C hybrid patterns, we performed a logistic regression using PCA‑reduced latent vectors to predict whether an A/C pattern remains amorphous (A) or transitions into an ordered state. Because the distinction between crystalline (C) and recrystallized (R) orientations becomes increasingly ambiguous at high temperature, we grouped these two low‑entropy ordered phases into a single outcome, denoted C$^\ast$. Treating C and R together for trajectory analysis provides a unified measure of ordering while preserving the contrast with the amorphous state. The model was trained on 290 A/C$^\ast$ samples, with an observed A$\rightarrow$C$^\ast$ transition ratio of 0.44. The fitted model yields a reaction coordinate
\[
\xi = w^\top z + b, \qquad b = -3.53,
\]
that separates dissolving from crystallizing A/C states. The corresponding commitment probability
\[
s = \frac{1}{1 + e^{-\xi}}
\]
varies smoothly across the A/C manifold (Figure S4 of the Supplemental Information), revealing a latent space gradient that predicts whether an A/C hybrid will anneal into C$^\ast$ or revert to A. The negative value of $b$ indicates the system’s preference for A states to remain A. This logistic-regression model functions analogously to a committor in transition-state theory: it assigns each A/C pattern a probability of crystallizing, and the surface where $s = 0.5$ defines a low-dimensional boundary separating dissolution (or retention) of A from C$^\ast$ pathways~\cite{chen2023discovering,talmazan2025coupling}. This committor-like analysis provides a quantitative, low-dimensional reaction coordinate for amorphous-to-crystalline transitions, uncovering nucleation precursors directly from diffraction data.

\begin{figure*}[t]
\centering
\includegraphics[width = 0.82\textwidth]{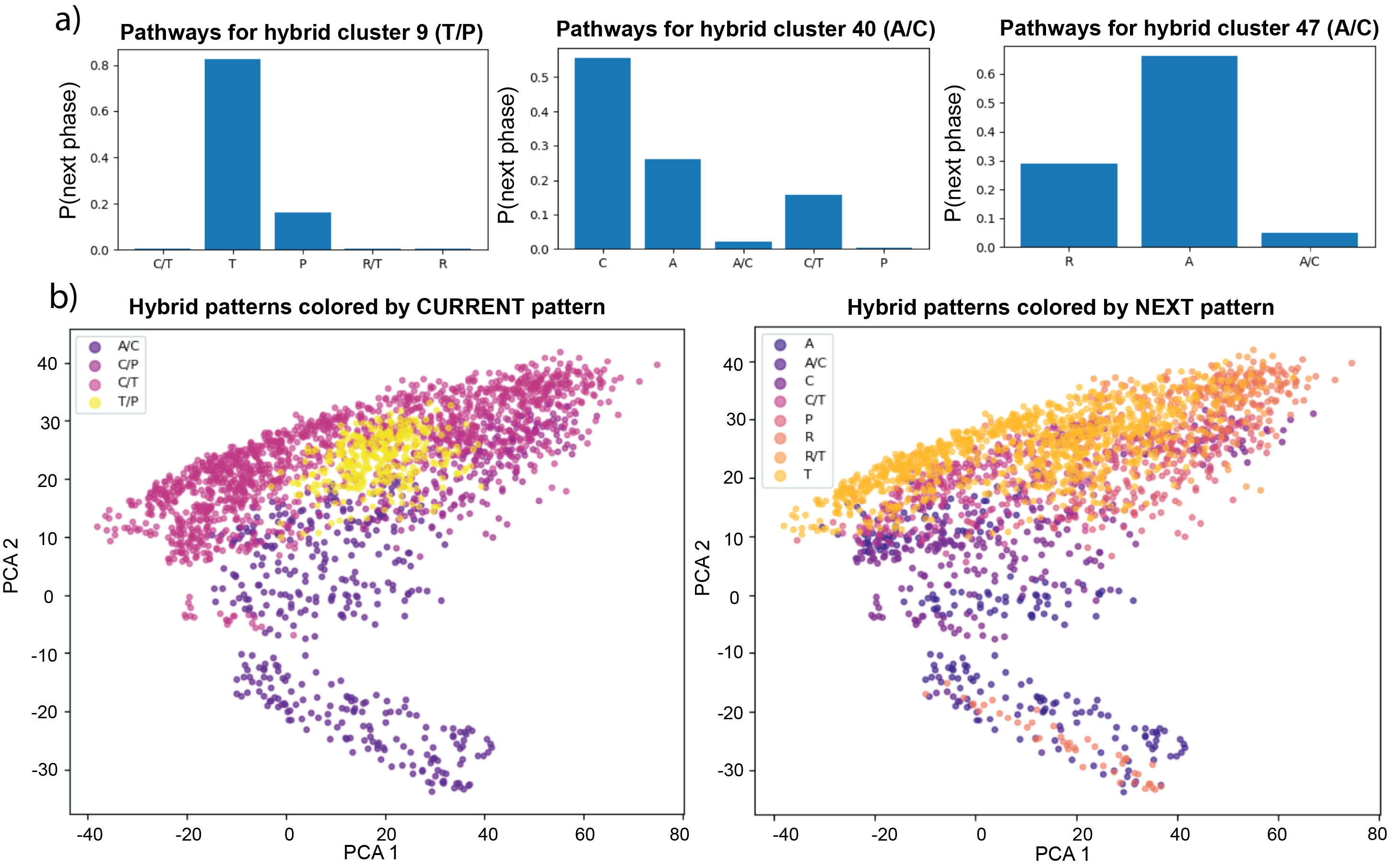}
 \caption{\small \textbf{Hybrid state intermediates and their role in recrystallization pathways.}
(a) Transition-destination probabilities for the hybrid clusters that exhibit measurable evolution across temperature steps. The T/P hybrid (cluster 9) and the two A/C hybrids (clusters~40 and~47) show distinct fates: cluster 40, which appears primarily at early temperatures, dissolves into A or anneals into R, while cluster 47 spans intermediate temperatures and exhibits broader distributions including transitions to C, T, and C/T at higher temperatures. The high-temperature C/T hybrid (cluster 9) overwhelmingly transitions to T, indicating that faulted crystalline regions act as direct precursors to twinning. Hybrid clusters that appear only at the final temperature (C/P and T/P) show no low-temperature precursors and transition primarily among P, R, and T, reflecting their terminal high-temperature character. (b) PCA projection of hybrid patterns colored by their current phase label (left) and by their next-phase outcome (right). A/C hybrids form multiple lobes corresponding to dissolution (A), crystallization (C), and twinning (T) pathways, while the C/T hybrids occupy a narrow ridge that flows almost exclusively to T. High-temperature hybrids (C/P and T/P) cluster tightly.}
  \label{fig:hybrid}
\end{figure*}

Although labeled as amorphous, a subset of diffraction patterns exhibits subtle structural signatures that precede recrystallization and are not discernible through conventional inspection. To identify these incipiently ordered regions, we analyzed all amorphous patterns that could be tracked to the subsequent temperature step. Across the full temperature series, a total of 17,731 amorphous A labeled patterns were evaluated, of which 7.87\% transitioned into an ordered state (C or R) at the next temperature step. These A$\rightarrow$C$^\ast$ events provide a direct window into the earliest stages of nucleation.

We first examined the position of these patterns in latent space. Amorphous patterns that later crystallize lie significantly farther from the amorphous cluster center than those that remain amorphous, indicating that the latent space representation encodes microstate stability: proximity to a macrostate’s center of mass reflects how strongly a pattern belongs to that basin. Patterns lying near the center of a macrostate basin exhibit high kinetic self-retention in the transition matrix analysis, whereas patterns displaced toward basin boundaries show elevated transition probabilities. The mean latent space distance for incipient amorphous patterns is 29.57, compared to 19.59 for stable amorphous patterns, where ``stable" refers to amorphous regions that remain amorphous at subsequent temperature steps. This $\sim$51\% increase indicates that the autoencoder captures subtle deviations from isotropic disorder in $k$-space that correlate with imminent ordering.

Stable amorphous patterns show weak, noise-level harmonic content, whereas incipient amorphous patterns exhibit clear symmetry enhancement. To probe the physical origin of these deviations, we computed the angular cross-correlation function (CCF) for each pattern and extracted harmonic amplitudes corresponding to two-fold, four-fold, and six-fold symmetry~\cite{liu2015interpretation,wochner2009xray}. Incipient amorphous patterns consistently show elevated harmonic content: the two-fold harmonic increases from 6.60 $\times$ 10$^4$ to 7.64 $\times$ 10$^4$, the four-fold harmonic from 4.36 $\times$ 10$^4$ to 1.40 $\times$ 10$^3$, and the six-fold harmonic from 3.75 $\times$ 10$^4$ to 1.63 $\times$ 10$^3$. Although small in absolute magnitude, these enhancements are highly systematic across hundreds of patterns, indicating the presence of short-range orientational order that precedes the emergence of Bragg peaks.

Figure~\ref{fig:amorphous} illustrates these effects: the mean CCF curves for stable and incipient amorphous patterns show a clear separation in angular correlations, and the corresponding symmetry spectra reveal elevated harmonic content in the latter~\cite{kennedy2025mapping,liu2013systematic}. The CCF curves for incipient amorphous patterns display pronounced even-fold symmetries, whereas stable amorphous patterns do not exhibit a unified trend in their angular structure. In PCA space, the commitment probability $q$ obtained from the logistic-regression classifier forms a continuous gradient within the amorphous manifold, with high-$q$ regions corresponding to patterns that subsequently crystallize~\cite{chen2023discovering}.

The amorphous matrix is therefore not structurally uniform. Instead, it contains localized regions of enhanced short-range order that act as precursors to crystallization. In Figure~\ref{fig:amorphous}b, all shown patterns have high $q$ scores and do recrystallize, yet the degree to which this ordering is visually apparent ranges from imperceptible to weak, partial Bragg-like features. latent space distance, reconstruction error, and CCF symmetry all provide consistent indicators of these incipiently ordered amorphous states, revealing the earliest detectable signatures of nucleation in the diffraction data.

\section*{Discussion}

A mechanistic view of recrystallization in ion irradiated GaAs is provided by latent space diffraction pattern representation and transition probability analysis. Understanding the transition pathways from the initial sandwiched amorphous structure (Figure~\ref{fig:overview}) to a stacking fault-driven heterogeneous crystalline system reveals the interplay between nucleation- and growth-limited thermal regimes, as well as the antagonism between an inherently low-density ion irradiated amorphous solid and the driving force toward energetically favorable recrystallization. Several themes emerge that clarify how a cubic semiconductor navigates a complex transformation landscape shaped by disorder, defect energetics, and temperature-dependent mobility. GaAs is a relatively simple material case, but some of the observations here can be generalized to more complex systems.

By establishing a rigorous latent space classification of all structural motifs, we were able to transition from static characterization to a dynamic, trajectory-based analysis of how these motifs evolve with temperature. This two-stage approach of structural identification followed by mechanistic pathway reconstruction enables a level of interpretability that is not achievable with conventional diffraction metrics or \textit{ex situ} recrystallization studies. To our knowledge, this work provides the first nanometer-resolved, \textit{in situ} observation of high-temperature recrystallization pathways in an amorphous solid, revealing structural intermediates that have not been accessible in ensemble-averaged measurements. Our approach complements phenomenological treatments such as Avrami analyses that yield effective rate constants and growth exponents~\cite{zhao2021tellurium,humphreys2004chapter} by resolving the microstructural origin of those kinetics. By identifying metastable hybrid, faulted, polycrystalline, and twinned states and quantifying the transitions between them, we directly connect macroscopic recrystallization behavior to the nanoscale pathways that govern it.

\begin{figure}[t]
\centering
\includegraphics[width=0.45\textwidth]{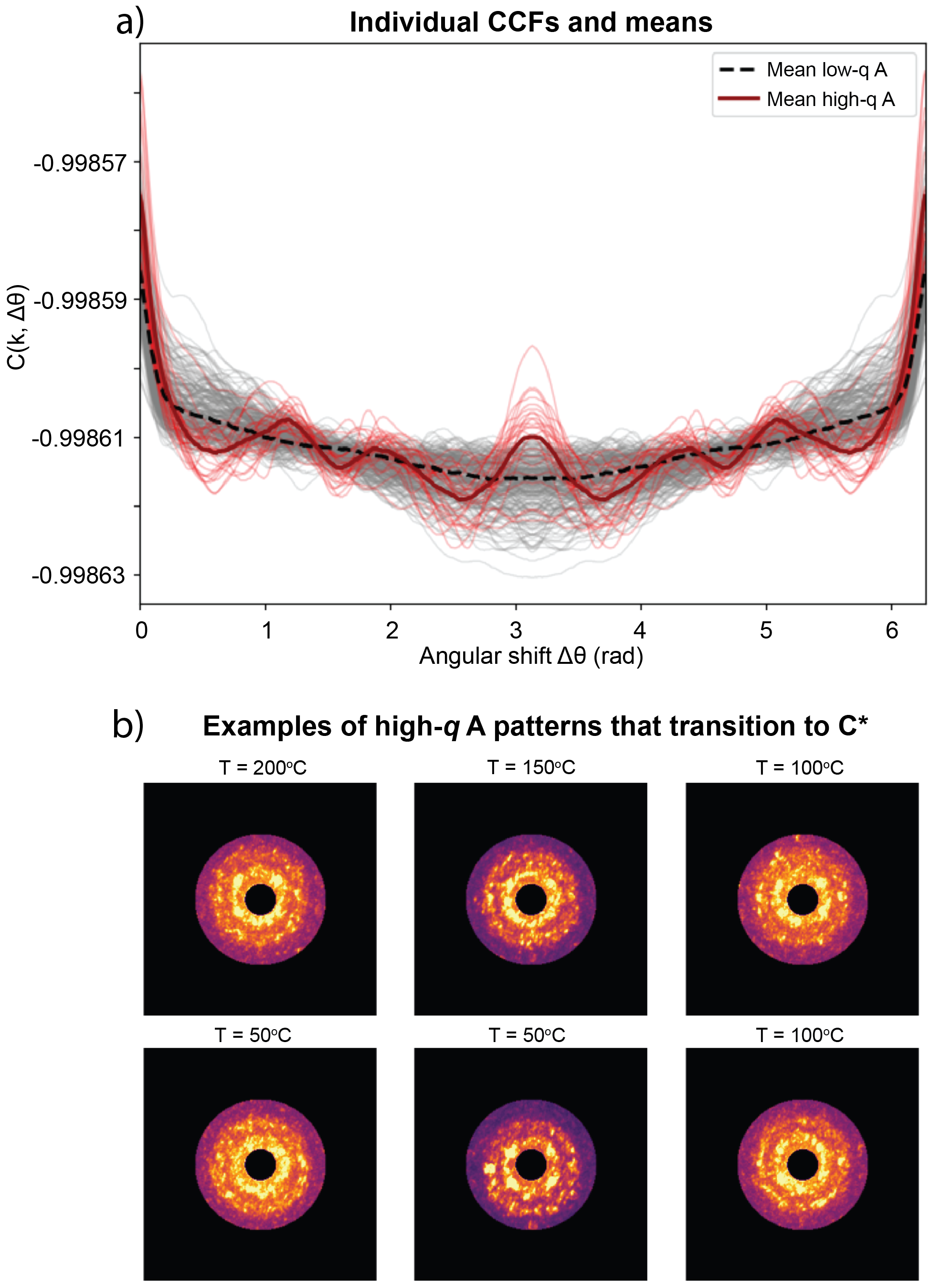}
\caption{\textbf{Angular cross-correlation scores and representative patterns of amorphous states with incipient crystallization.} (a) Individual angular cross-correlation functions (CCFs) for amorphous patterns that remain amorphous (low-$q$, gray) and those that subsequently crystallize into C$^\ast$ (high-$q$, red). The mean curves (black dashed for low-$q$ A; solid red for high-$q$ A) show systematic angular modulation in the high-$q$ population, indicating the emergence of short-range ordering prior to crystallization. (b) Representative high-$q$ amorphous patterns that transition to C$^\ast$, shown at their respective temperatures. These patterns exhibit subtle weak angular structure despite lacking or minimal Bragg peaks, illustrating the earliest detectable signatures of ordering within the amorphous matrix.}
\label{fig:amorphous}
\end{figure}

At low temperature, the system is growth-dominated, with the A and C phases dominating and remaining highly stable. As qualitatively described in Kennedy \textit{et al.}~\cite{kennedy2026critical}, there is minimal phase transformation in this regime, and what does occur is slow epitaxial recrystallization from the top and bottom crystalline layers. At high temperature, the system enters a far more dynamic and structurally volatile regime. The delicate reorganization established at lower temperatures is quickly overtaken as crystalline regions nucleate and expand to their critical size and beyond. Ion irradiated GaAs forms a low-density amorphous solid with significant free volume and disrupted tetrahedral bonding~\cite{cui2019what,light1969density}. Radiation-induced density reduction and defect accumulation destabilize the amorphous network, and once atomic mobility increases at high temperature, multiple misoriented crystalline embryos grow concurrently; lacking sufficient time or mobility to coalesce into a single orientation, the system transitions into polycrystalline configurations. This structural deficit raises the nucleation barrier and suppresses medium-range order, explaining the strong stability of the amorphous basin at low temperature. The low density associated with ion irradiation further hinder the formation of a uniform crystalline network. Consequently, the proportion of ordered patterns - C, R, P, and T - increases rapidly, effectively filling in amorphous regions with more energetically favorable configurations and disregarding any underlying structural templating, with the exception of the twins which, without fault (pun intended), follow the \{111\} planes.

Similarly, the hybrid state manifold splits into two distinct regimes. At low temperature, A/C hybrids correspond to partially ordered regions situated at the amorphous-crystalline interface, and their transitions between A and A/C primarily track fluctuations in the interface position rather than nucleation dynamics. The earliest A/C hybrid (cluster 47, mean temperature 0.53) predominantly dissolves into the amorphous phase (A = 0.66), with a minority transitioning into recrystallized order (R = 0.29), while the intermediate-temperature A/C hybrid (cluster 40) exhibits a strong bias toward crystallization (C = 0.56) with additional flux toward fault formation (C/T = 0.16) and dissolution (A = 0.26). Their transition probabilities, latent space geometry, and symmetry signatures all indicate that these states represent genuine structural precursors rather than noise or mixed labels. 

At high temperature, however, the hybrid manifold reorganizes into a metastable basin composed of C/T, C/P, T/P, and R/T intermediates. These states no longer dissolve; instead, they act as gateways that direct the system toward specific high-temperature outcomes. The hybrid manifold therefore serves as the connective tissue of the recrystallization landscape, encoding the short-lived but mechanistically essential intermediates that mediate transitions between amorphous, crystalline, polycrystalline, and twinned configurations. This partition reflects a fundamental shift in the underlying free energy landscape: the system transitions from one in which ordering is limited by nucleation barriers to one in which defect mobility, stacking fault formation, and grain coalescence dominate the kinetics.

Within this reorganized landscape, twinning emerges as the dominant terminal state. Analyzing the transition probabilities reveals that twinned configurations act as the primary sink for transitions originating in crystalline, hybrid, and polycrystalline regions. This behavior is consistent with the low stacking fault energy of GaAs and the well-established tendency of III-V semiconductors to accommodate strain through \{111\} faulting. At elevated temperature, the energetic penalty for forming intrinsic or extrinsic stacking faults becomes small relative to the driving force for structural recovery, making nanotwin formation an efficient pathway for relieving local strain while enabling rapid reordering. The strong transitions from C/T hybrids to T (0.46-0.57) and from T/P hybrids to T (0.83) observed here align with prior observations in Ge and Si, where grain coalescence and boundary migration generate dense twin networks during high-temperature annealing. Our results provide a quantitative, state-resolved view of this process, showing how faulted intermediates funnel the system toward twinning as the preferred structural endpoint. The observed transition pathways also provide evidence regarding the longstanding question of annealing-twin formation. In contrast to models invoking the impingement of pre-twinned nuclei, the dominant trajectories pass through faulted crystalline and polycrystalline intermediates before reaching the twinned state~\cite{lin2015observation}. In irradiated GaAs, twins therefore emerge primarily as a consequence of stacking-fault propagation and grain coarsening during the final stages of recrystallization.

The transition matrix and trajectory analysis also clarifies the role of polycrystalline regions in mediating high-temperature recrystallization. The C/P hybrid (cluster 37) transitions overwhelmingly to P (0.81), while T/P hybrids split between sharpening into twins (T = 0.83) and coarsening into polycrystalline grains (P = 0.16). These behaviors indicate that small, defective grains do not necessarily persist as stable microstructural features; instead, they may coalesce, rotate, or migrate to form larger ordered domains. As temperature increases, boundary migration becomes energetically favorable, enabling grains to merge and defects to reorganize. The resulting structures - either defect-free recrystallized grains or faulted twins - represent low-energy configurations that minimize both elastic and interfacial energy. The terminal nature of the R/T hybrid (cluster 14), which exhibits no outgoing transitions, further supports this picture: once a grain achieves a sufficiently low defect density and adopts a faulted configuration, it becomes energetically isolated from the surrounding transformation dynamics.

The latent space representation provides additional insight into the microscopic origins of these pathways. The autoencoder preserves diffuse scattering, peak asymmetry, and weak symmetry signatures that are essential for distinguishing intermediate structural states (Figure S1 of the Supplemental Information). These features are essential for distinguishing intermediate states whose differences lie in subtle variations of disorder and symmetry. The resulting latent manifold is not an abstract embedding but a physically interpretable space in which distances and directions correspond to meaningful structural changes. The separation of A/C, C/T, and T/P hybrids, and the smooth gradients observed in the logistic-regression reaction coordinate for A$\rightarrow$C$^\ast$, demonstrate that the latent space encodes signatures of the underlying physics required to predict disorder-to-order transitions.

To ensure clean phase boundaries, we intentionally over‑clustered the latent space by choosing 50 clusters in the k‑means algorithm, producing multiple clusters within each physical phase. This strategy reduces boundary smearing and allows hybrid states to emerge naturally as clusters that lie between major basins. After clustering, we grouped clusters into physical classes (A, C, R, P, T, and H) based on diffraction signatures and latent space geometry. Of the 50 clusters, eight correspond to amorphous, eleven to crystalline, seven to recrystallized, eight to polycrystalline, six to twinned, and the remaining nine to hybrid intermediates. This over-clustering approach is essential for resolving subtle structural motifs that would be obscured in a coarse-grained classification and provides the granularity required to identify distinct hybrid pathways.

The division between low- and high-temperature regimes aligns with known behavior of GaAs, which has a reported recrystallization temperature between 300-400\textdegree{}C in bulk and as low as 177\textdegree{}C in thin-film geometries. Below $\sim$250\textdegree{}C, the amorphous matrix retains medium-range order inherited from the crystalline parent but lacks the mobility required for defect annihilation or interface motion. This regime is dominated by subcritical nuclei, paracrystalline motifs, and density fluctuations that either anneal or dissolve depending on local strain and disorder. Above this threshold, increased atomic mobility enables stacking fault formation, grain coalescence, and rapid interface motion, producing the growth-limited (nucleation-dominated) regime observed in prior \textit{in situ} studies. The latent space and transition probability analyses capture this transition directly from the diffraction data, without requiring predefined structural models or assumptions about the nature of intermediate states. We note that limited transformation at low temperature may also reflect suppressed mobility more generally; however, the latent space trajectories and hybrid state distributions are most consistent with a growth-dominated to nucleation-dominated crossover.

Our analysis of transition probabilities is inspired by the use of Markov State Models in molecular simulation~\cite{bowman2013introduction}.
These models track transitions between clusters in simulated system's phase space to construct a matrix of transition probabilities.
We do not refer to our analysis as a Markov state model to avoid overinterpretation of our results.    In Markov state models, the matrix of transition probabilities is taken to be an approximation of the the system's propagator~\cite{Sarich2014}.
In contrast, our transition probability matrices are taken over an inherently time-dependent dynamics as sampled points are taken between different temperatures.  Moreover, they ignore spatial correlation between areas of the sample.  

Rather than as a rigorous approximation of the system's propagator, our matrices the matrices should be interpreted as summarizing the dominant, coarse-grained pathways within each temperature regime, capturing the directionality and relative likelihood of structural evolution. Our results are therefore best viewed as a trajectory-informed map of the transformation landscape-one that reveals how structural motifs reorganize, merge, and funnel into terminal states as temperature increases.

The ability to predict phase change before it is complete motivated an analysis of A-classified patterns that transition into an ordered state (C or C$^\ast$) at the next temperature step. Only a small subset of amorphous patterns, 7.87\%, exhibit this transition: at low temperature the amorphous phase is stable, and above 250\textdegree{}C almost no amorphous material remains. Thus, nearly all A$\rightarrow$C$^\ast$ events occur at the 250\textdegree{}C step. Identifying amorphous patterns with weak but systematic symmetry signatures provides structural evidence for incipient ordering within the amorphous matrix. Representative examples are shown in Figure~\ref{fig:amorphous}b, where mild crystallinity is apparent to varying degrees, including cases where it is not intuitively visible. These patterns exhibit enhanced harmonic content in their angular cross-correlation functions and occupy distinct regions of latent space, indicating that the amorphous phase is not structurally uniform. These incipiently ordered amorphous regions resemble the medium-range order fluctuations predicted in paracrystalline models of amorphous semiconductors and thin films, where local bond-angle correlations and density variations create structurally privileged sites for nucleation. The latent space distance, CCF symmetry, and logistic-regression committor are in agreement, demonstrating that the earliest signatures of ordering are encoded in subtle angular modulations long before Bragg peaks appear.

As shown in Figure~\ref{fig:amorphous}a, low-$q$ amorphous patterns display variable correlation scores across $k$-space, consistent with isotropic disorder. In contrast, high-$q$ patterns exhibit substantially stronger harmonic content, dominated by 6-fold symmetry. This observation is striking because the C$^\ast$ phase itself exhibits primarily 2-fold symmetry when zincblende is viewed along [100], as demonstrated in single-crystal GaAs by Kennedy \textit{et al.}~\cite{kennedy2026critical}, where the 2-fold component overwhelms all other $n$-fold contributions. However, in the incipient amorphous state, while the signal is weak, the dominant symmetry is 6-fold (and, to a lesser extent, its 3-fold base harmonic at 2$\pi$/3), not 2-fold. This implies that as the amorphous matrix reorganizes toward an ordered configuration, it initially develops tetrahedrally coordinated short-range order with 6-fold angular modulation, which is consistent with the local bonding geometry of zincblende, before subsequently reorienting to match the [100] orientation of the surrounding crystalline regions. In other words, while remnant structural memory and paracrystallinity may guide reordering in some cases (e.g., lower displacements per atom near the surface or free surfaces abutting ordered regions), the amorphous precursor does not necessarily “know” the final crystallographic orientation; it first forms locally ordered motifs with the symmetry of tetrahedral coordination and only later snaps into registry with the neighboring C or R domains. The presence of paracrystalline motifs within the amorphous matrix is consistent with paracrystalline models of amorphous semiconductors, in which locally ordered fragments coexist with topological disorder. These motifs act as the structural seeds for the A/C hybrid states observed here. Significantly, even without long-range order, the amorphous incipient structure expresses the correct local symmetry of zincblende, albeit at an orientation that is later rapidly corrected.

These results indicate that the amorphous matrix contains localized regions of short-range order, paracrystalline clusters, and density-modulated motifs that act as subcritical nuclei. Such features have been proposed in simulations, inferred from calorimetry, and qualitatively observed in TEM, but they are rarely resolved directly in diffraction data. Their detection here demonstrates the sensitivity of the latent space and CCF-based approach to subtle structural signatures that precede crystallization, providing a quantitative, state-resolved view of the earliest stages of nucleation.

The logistic-regression model provides an additional window into these precursors. Because the reaction coordinate $\xi = w^\top z + b$ is a linear function of the PCA-reduced latent vector, the learned weight vector $w$ identifies the latent space directions most strongly associated with imminent ordering. By projecting $w$ back through the decoder, one could in principle visualize the $k$-space features that most strongly influence the crystallization probability-highlighting the angular sectors, diffuse-scattering modulations, or weak symmetry components that the model interprets as nucleation signatures. This offers a path toward physically interpretable “saliency maps” for amorphous-to-crystalline transitions, linking the statistical structure of the latent space to real-space or reciprocal-space motifs that precede ordering.

The use of unsupervised clustering to identify 50 microstructural states is a deliberate choice to balance interpretability with resolution. This redundancy and granularity allow the latent space to capture the continuous nature of structural evolution while enabling the trajectory analysis to resolve distinct intermediates with meaningful transition statistics. The emergence of physically interpretable clusters (A/C, C/T, C/P, T/P, R/T) validates this choice and highlights the ability of data-driven methods to uncover mechanistic structure in complex transformation pathways.

The trajectory-based framework is complementary to established microstructural modeling approaches such as Potts~\cite{rollett1989computer,anderson1984computer}, phase-field~\cite{tikare1998comparison}, and atomistic simulations~\cite{jackson1995monte}. Whereas these methods prescribe local interaction rules and evolve a simulated microstructure forward in time, our approach reconstructs the transformation landscape directly from STEM experiment, identifying metastable states, transition probabilities, and reaction coordinates without introducing phenomenological kinetic parameters. It provides a quantitative bridge between experiment and simulation where experimentally measured transition rates and state occupancies can be used to constrain or calibrate mesoscale models, while the observed intermediate structures offer physically grounded targets for model validation. Additionally, the ability to infer microstructural pathways directly from diffraction data suggests a route toward data-driven simulations in which experimentally determined state networks inform the construction of predictive phase-field or Potts-type descriptions of phase transformations.

These results demonstrate that a latent space- and MSM-based analysis can resolve the multipathway, defect-mediated nature of recrystallization in GaAs. The framework captures metastability, stacking-fault energetics, grain-boundary migration, and temperature-dependent reorganizations of the transformation landscape, providing a quantitative and mechanistic view of structural recovery in a covalent cubic semiconductor. More broadly, this approach offers a generalizable route for mapping phase-transformation pathways in disordered and partially ordered materials, where conventional descriptors fail to capture the richness of the underlying dynamics.

\section*{Conclusions}

By integrating a convolutional autoencoder with latent‑space clustering and transition matrix analysis, we establish a two‑stage framework for understanding recrystallization in ion‑irradiated GaAs: latent‑space structural classification first, followed by trajectory-based dynamical analysis to extract mechanistic pathways. This workflow reveals a bifurcation in the transformation landscape: persistent amorphous and crystalline basins dominate at low temperature, whereas high‑temperature evolution reorganizes into hybrid‑mediated pathways that funnel into polycrystalline, recrystallized, and ultimately twinned end states. The latent representation also exposes amorphous patterns with weak but systematic symmetry signatures, incipient ordering that is invisible to conventional descriptors, demonstrating that the amorphous matrix encodes hidden structural predispositions that shape nucleation. Together, these results show that combining latent space modeling with nanometer‑resolved trajectory analysis provides a physics‑informed route for mapping transformation pathways directly from 4D‑STEM data.

Future extensions involve more advanced computational treatments of the dynamics, the use of faster detectors to minimize time‑dependent evolution during acquisition, and eventually an end‑to‑end automated workflow. However, the present approach is deliberately built around defined inputs, interpretable checks, and physics‑guided constraints, and already provides a robust methodology with intentional user input. More broadly, this work offers a route for quantifying \textit{in situ} recrystallization dynamics and for measuring embedded structural information within amorphous materials, enabling new ways to understand how disorder, symmetry, and metastability govern transformation pathways in complex solids.

\section*{Methods}

\subsection*{Sample preparation}

The procedures for GaAs irradiation, cross-section preparation, and 4D-STEM acquisition are described in Kennedy \textit{et al} \cite{kennedy2026critical}. Briefly, [001]-oriented GaAs wafers were irradiated with 400~keV Ne$^{++}$ ions to a fluence of $2 \times 10^{16}$~ions/cm$^{2}$ at the Ion Beam Materials Laboratory (Los Alamos National Laboratory). Cross-sectional TEM lamellae were prepared using a Helios G4 UX dual-beam SEM/FIB and final-thinned to approximately 60-80 nm on a Protochips thermoelectric MEMS chip. After several weeks at ambient conditions, partial spontaneous recrystallization was observed near the free surface and at the crystalline-amorphous interface.

\subsection*{4D-STEM data acquisition}

4D-STEM experiments were performed on an image-corrected FEI Titan operated at 300~keV using a OneView 4k CCD detector. Data were collected in microprobe mode with a 770~mm camera length and a 50~$\mu$m condenser aperture, yielding a convergence angle of $\sim$0.76~mrad and a probe diameter of $\sim$2.6~nm. A 5~nm step size was used to provide partial probe overlap.

\textit{In situ} annealing was conducted under high vacuum using the Protochips heating chip. The sample was heated in 50\textdegree{}C increments from room temperature to 400\textdegree{}C. At each temperature, the lamella was stabilized for approximately one minute before acquiring full 4D-STEM datasets from Regions~1 and 2, with each acquisition lasting roughly 12 minutes. This procedure was repeated sequentially at 25\textdegree{}C, 50\textdegree{}C, 100\textdegree{}C, and so on up to 400\textdegree{}C.

\subsection*{4D-STEM dataset alignment}
4D-STEM data were loaded and preliminarily analyzed using the \texttt{py4DSTEM} Python package~\cite{savitzky2021py4dstem}, \texttt{scikit-learn}~\cite{scikitlearn}, and custom scripts. 

To enable direct comparison of structural evolution across temperature, the eight temperature-series 4D-STEM datasets were aligned onto a common real-space scan grid. Bright-field (BF) images acquired simultaneously with the diffraction patterns were first symmetrically cropped to a fixed $64 \times 80$ scan geometry. Real-space alignment was performed by identifying corresponding fiducial features across temperature steps. These fiducials included (i) the advancing recrystallization fronts and (ii) a small carbon contamination spot produced by briefly parking the electron beam. For each temperature, the displacement between fiducial pairs was converted from the lower-left coordinate system of the acquisition software into image row-column coordinates, and the mean translation was applied to the BF image. All BF images were then placed onto a common canvas, and the global region of overlap across all temperatures was extracted.

The same translations were applied to the diffraction patterns. Raw patterns were first reshaped into $(N_y, N_x, q_y, q_x)$ arrays matching the BF scan geometry, and pattern indices were placed onto the BF-aligned canvas. Only scan positions present at all temperature steps were retained, ensuring that each spatial location had a complete temperature trajectory. The resulting aligned 4D-STEM datasets were written to HDF5 files for downstream analysis.

Diffraction-space alignment was verified by inspecting representative diffraction patterns in $k$-space across temperatures. The positions of Bragg reflections and diffuse scattering features were consistent after real-space alignment, and no additional $k$-space registration was required. Thus, the final alignment was performed primarily in real space, with $k$-space checks confirming structural consistency.

\subsection*{Diffraction pattern pre-processing}

Diffraction patterns from the aligned 4D-STEM datasets were flattened into a list of individual patterns, each associated with its scan position $(y,x)$, temperature index, and a unique global identifier. Preprocessing was performed to reduce intensity variations unrelated to structural evolution and to ensure consistent centering across the full dataset.

First, the direct beam was removed by masking a circular region of radius 17 pixels around the measured probe center $(q_{x0}, q_{y0})$. The probe center was then shifted to the center of the diffraction frame using a subpixel translation. An optional outer mask was applied to remove high-angle scattering beyond a radius of 72 pixels, reducing noise from detector edges and enhancing sensitivity to medium-range order.

Each pattern was normalized by its 99th-percentile intensity to suppress outlier pixels and to ensure consistent dynamic range across temperatures. Intensities were clipped to the range $[0,1]$ to stabilize training of the autoencoder. This preprocessing produced a set of masked, centered, and normalized diffraction patterns of size $256 \times 256$, each associated with its spatial coordinates, temperature, and global ID.

\subsection*{Latent space learning with a convolutional autoencoder}

A convolutional autoencoder (CAE) was used to obtain a compact latent space representation of the diffraction patterns. The encoder consisted of four convolutional blocks, each using a $3\times 3$ kernel, stride~2, and padding~1, which progressively increased the number of feature channels from 1~→~16~→~32~→~64~→~128 while downsampling the spatial resolution by a factor of two at each layer. This reduced the $256\times 256$ input pattern to a $(128 \times 16 \times 16)$ encoded feature map. The encoded volume was flattened and projected into a 218-dimensional latent vector using a fully connected layer. Several reconstruction losses (L1, MSE/RMSE) were evaluated and L1 was found to produce reconstructions that better preserved diffuse scattering and weak symmetry features, despite slightly higher numerical error under RMSE. Example reconstructions are provided in Figure S1 of the Supplemental Information.

The decoder mirrored the encoder using transposed convolutions with the same kernel size, stride, and padding, expanding the feature maps from 128~→~64~→~32~→~16~→~1 channels and reconstructing the full $256\times 256$ diffraction pattern. A final sigmoid activation ensured that reconstructed intensities remained within the normalized $[0,1]$ range.

The dataset was randomly split into 75\% training and 25\% validation sets. The CAE was trained for 100~epochs using the Adam optimizer (learning rate $10^{-3}$) and an L1 reconstruction loss (Figure S1 of the Supplemental Information). A stepwise learning-rate scheduler (factor 0.5 every 10~epochs) was applied. Training and validation losses decreased smoothly, and reconstructed patterns preserved the main diffuse and Bragg-scattering features, indicating that the latent vectors captured the essential structural information.

After training, latent vectors were extracted for all patterns in the dataset. The global IDs were used to restore the original spatial and temperature ordering, producing a latent space array aligned with the full 4D-STEM experiment.

\subsection*{Dimensionality reduction and clustering}

To identify recurring microstructural states, the CAE latent vectors were analyzed using principal component analysis (PCA) followed by $k$-means clustering. After CAE training, the full dataset was recombined and split 90/10 into training and test sets. PCA was fit on the 90\% training subset and reduced the 218-dimensional latent vectors to 64 principal components. The same PCA transformation was applied to the 10\% test subset. The first two PCA components were used to visualize and interpret the latent space structure.

$k$-means clustering was performed on the PCA-transformed training data. A total of 50 clusters were used, chosen as a conservative value that intentionally over-partitions the latent space. This duplication of similar microstates ensured that the resulting clusters formed clear, well-separated boundaries between the structural phases. Cluster labels were assigned to the test data using the trained model. Clusters were manually grouped into physically interpretable phases (amorphous, polycrystalline, crystalline, twinned, polycrystalline) based on representative diffraction patterns from each cluster. Each pattern therefore had an associated microstate label, phase label, spatial coordinate, and temperature index.

Cluster labels were mapped back onto the $(y,x)$ scan grid for each temperature, producing a temperature-resolved phase map for the entire dataset.

\subsection*{Quantify transition paths prediction}

To quantify recrystallization pathways, microstate trajectories were constructed for each scan position by tracking its cluster label across temperature steps. A transition count matrix was computed by comparing microstate identities at temperature $T_i$ and $T_{i+1}$ for all spatial positions. Row-normalization yielded a transition probability matrix $P$ describing the likelihood of transitioning between microstates as temperature increased.

Phase-level transition matrices were obtained by coarse-graining microstates into their assigned phases. These matrices revealed dominant pathways, including (i) a low-temperature amorphous-to-crystalline transition and (ii) two high-temperature pathways, one direct and one passing through an intermediate twinned state. To focus on robust transformation pathways, we visualized only transitions with probabilities $\geq{}3\%$ and at least 10 observed events; lower-probability transitions are retained in the numerical model but are not interpreted mechanistically or represented in Figure~\ref{fig:transitions}.

For prediction, each test-set pattern was mapped into PCA space, assigned a cluster label, and its next-state probabilities were computed using the learned transition matrix. This enabled prediction of the most likely structural evolution pathway for any given diffraction pattern and temperature.

\subsection*{Hybrid state identification and pathway analysis}

Hybrid states were identified by first clustering the PCA-reduced latent vectors using $k$-means (50 clusters). Clusters were manually assigned to structural phases based on representative diffraction patterns. Clusters whose labels contained two phase identifiers (e.g., A/C, C/T, C/P, R/T) were designated as hybrid microstates. These hybrid clusters occupy boundary regions in latent space and represent partially ordered or defect-mediated configurations.

To quantify their mechanistic role during recrystallization, we tracked each hybrid-labeled scan position across consecutive temperature steps. For each pixel with a hybrid label at temperature $T_i$, the corresponding label at $T_{i+1}$ was retrieved using its $(y,x)$ coordinate. This produced a set of hybrid to next-phase transitions for all hybrid clusters. Transition probabilities were computed by normalizing the counts of next-phase outcomes for each hybrid cluster.

Temperature distributions for each hybrid cluster were obtained by collecting the temperature indices of all hybrid-labeled patterns. These distributions were used to determine whether a hybrid cluster appeared exclusively at low temperature, high temperature, or across the full series. To visualize the geometric structure of hybrid evolution, PCA coordinates of hybrid patterns were colored by either their current phase label or their next-phase outcome. These “fate maps” revealed low-dimensional directions in latent space associated with dissolution, crystallization, twinning, or grain coalescence. Together, these analyses provided a coarse-grained description of metastable intermediates and their preferred transformation pathways.

\subsection*{Analysis of amorphous precursors to crystallization}

To identify amorphous patterns that act as precursors to crystallization, we first isolated all patterns labeled as amorphous (A) in the phase map. For each amorphous pattern at temperature $T_i$, the corresponding pattern at $T_{i+1}$ was retrieved using its $(y,x)$ coordinate. Amorphous patterns whose next-phase label was crystalline (C) or recrystallized (R) were designated as incipient amorphous states (A $\rightarrow$ C$^\ast$), while those that remained amorphous formed the stable A population.

latent space distances were computed by measuring the Euclidean distance between each amorphous latent vector and the centroid of all A-labeled latent vectors. Reconstruction errors were obtained by passing each pattern through the trained CAE and computing the mean absolute difference between the input and reconstructed diffraction pattern. These metrics quantified deviations from the amorphous manifold.

Angular cross-correlation functions (CCFs) were computed for each pattern by radially integrating the diffraction intensity and evaluating the angular autocorrelation $C(k,\Delta\theta)$ at fixed scattering vectors. The CCF is a function of the probe position $\mathbf{r} = (x, y)$ and the scattering vector $k$:
\begin{align*}
C(\mathbf{r},k,\Delta) = \frac{\langle I(k,\varphi)I(k,\varphi+\Delta) \rangle - \langle I(k,\varphi)\rangle^2}{\langle I(k,\varphi)\rangle^2}
\end{align*}
where $I(k,\varphi)$ is the diffracted intensity at a given scattering vector $k$, averaged over the azimuthal angle $\varphi$~\cite{liu2015interpretation}.

The CCFs shown in Figure~\ref{fig:amorphous} were calculated using a narrow band of $k$-space corresponding to the region where the amorphous ring and the \{111\} Bragg peaks appear to selectively capture rotational correlations associated with the amorphous-to-crystalline transition in that specific portion of $k$-space (4.1 - 8.4 mrad). Harmonic amplitudes were extracted by Fourier transforming the CCF with respect to $\Delta\theta$, yielding symmetry components at $n=2,4,6$. Mean CCFs and harmonic spectra were computed separately for stable and incipient amorphous populations. Representative patterns were selected by sampling high-$q$ and low-$q$ subsets.

\section*{Acknowledgments}

This project is primarily supported by the Eric and Wendy Schmidt AI in Science Postdoctoral Fellowship, a program of Schmidt Sciences, LLC. E.R.K. and K. J. are both supported by Schmidt Sciences, LLC. This work used SDSC Expanse GPU at the San Diego Supercomputer Center (SDSC) through allocation  MAT260007 from the Advanced Cyberinfrastructure Coordination Ecosystem: Services \& Support (ACCESS) program, which is supported by U.S. National Science Foundation grants \#2138259, \#2138286, \#2138307, \#2137603, and \#2138296~\cite{boerner2023access}. This work was also supported by LDRD project number 20240033DR. This work was performed, in part, at the Center for Integrated Nanotechnologies, an Office of Science User Facility operated for the U.S. Department of Energy (DOE) Office of Science. Los Alamos National Laboratory is operated by Triad National Security, LLC, for the National Nuclear Security Administration of U.S. Department of Energy (Contract No. 89233218CNA000001).

\section*{Data Availability}
All data needed to evaluate the conclusions in the paper are present in the paper and/or the Supplemental Materials. The raw 4D-STEM data are openly available at 10.5281/zenodo.17416428. 

\section*{Author Contributions}
Ellis Rae Kennedy: Writing - review \& editing, Writing - original draft, Visualization, Validation, Methodology, Investigation, Funding acquisition, Formal analysis, Data curation, Conceptualization. Kwanghwi Je: Writing - review \& editing, Writing - original draft, Visualization, Methodology, Investigation. Erik Thiede: Writing - review \& editing, Writing - original draft, Validation, Supervision, Resources, Methodology, Funding acquisition, Conceptualization.

\section*{Competing Interests}
The authors declare no competing interests.

\bibliography{main}

\end{document}


\vspace*{-1.5cm} 
\begin{center}
    \Large
    \textbf{Supplemental Information for}\\[0.5em]
    ``Mapping recrystallization trajectories in GaAs using latent space diffraction analysis''
\end{center}
\vspace{0.5cm}

\begin{figure*}[!ht]
\centering
\includegraphics[width=0.8\textwidth]{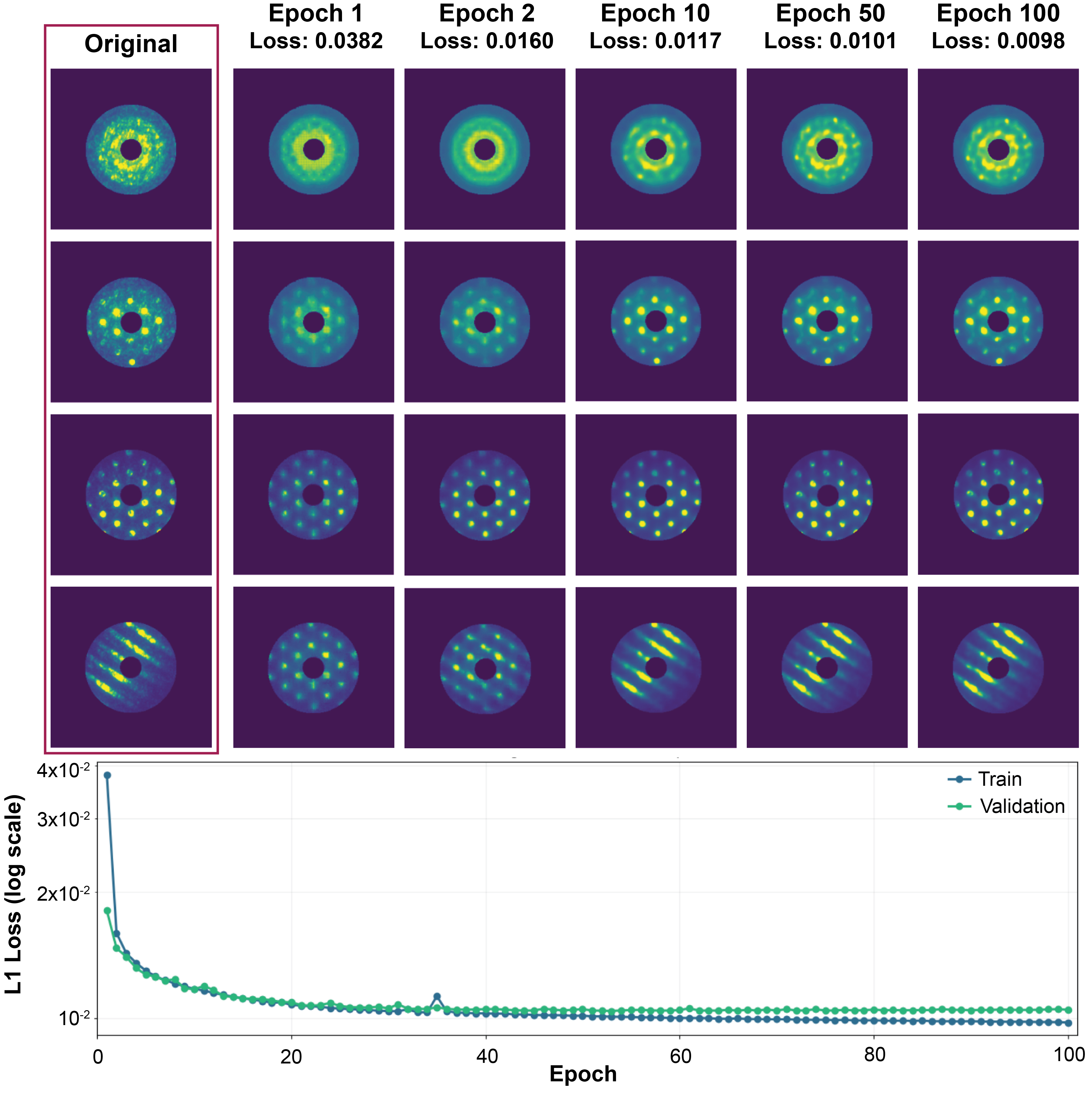}
\caption{Autoencoder reconstruction fidelity and training convergence plot.
Representative reconstructions (left) and their evolution over training epochs (right) demonstrate that the convolutional autoencoder learns a physically meaningful latent representation. Early epochs capture only coarse diffuse features, while later epochs reproduce Bragg peak positions, symmetry, and diffuse-scattering structure with high fidelity. The training and validation losses converge smoothly, indicating stable learning and preventing overfitting. Based on the loss curves and validated by the reconstructions, above roughly 50 epochs there is negligible improvement in feature reconstruction.}
\label{fig:cae}
\end{figure*}

\begin{figure*}[!ht]
\centering
\includegraphics[width=0.8\textwidth]{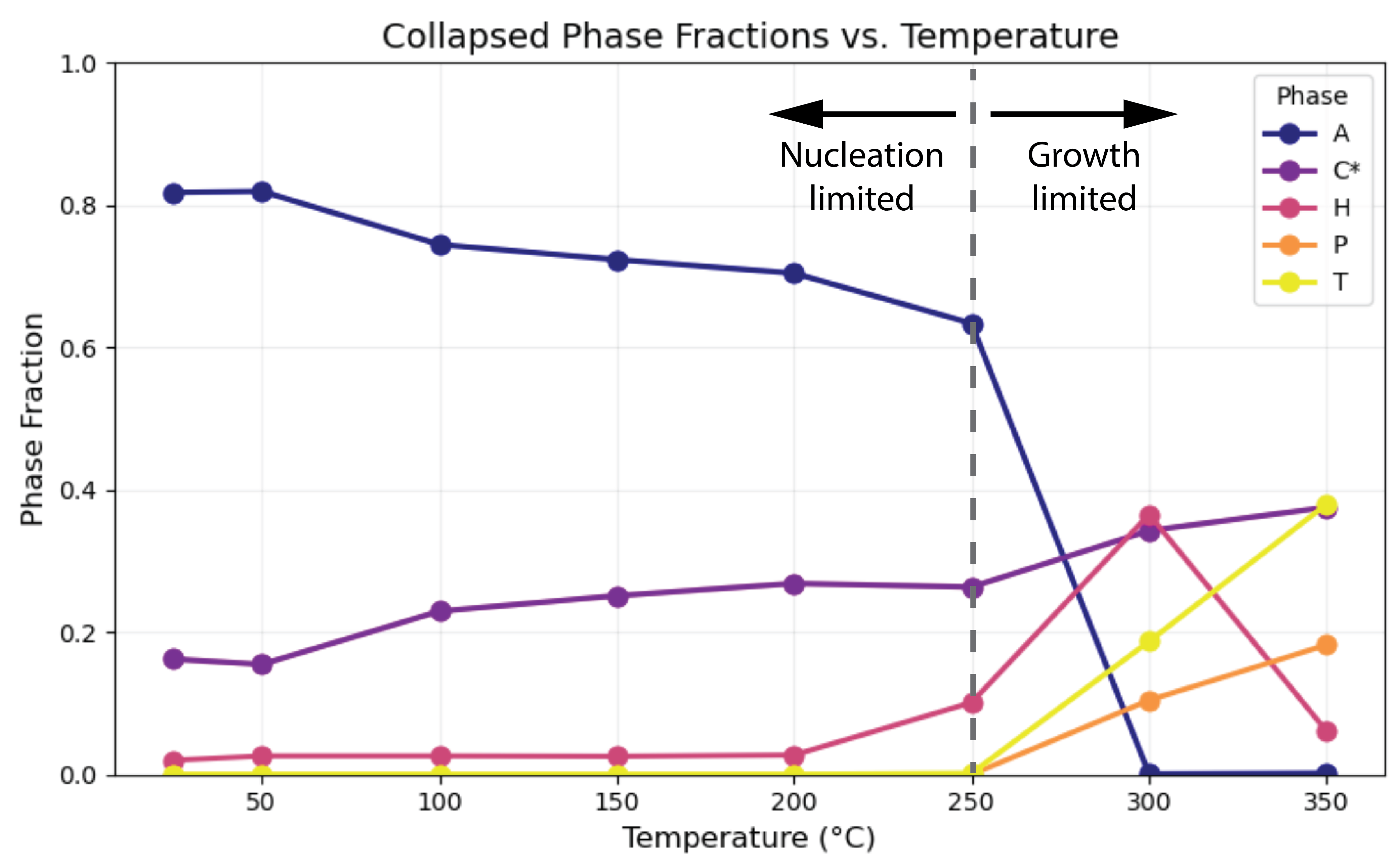}
\caption{\label{fig:figs1} Fraction of counts for each phase, excluding hybrid classifications, as a function of annealing temperature. The system shifts from nucleation limited to growth limited recrystallization at approximately 250\textdegree{}C. Accordingly, there is a dramatic decrease in amorphous phase patterns above this temperature.}
\end{figure*}

\begin{figure*}[!ht]
\centering
\includegraphics[width=0.8\textwidth]{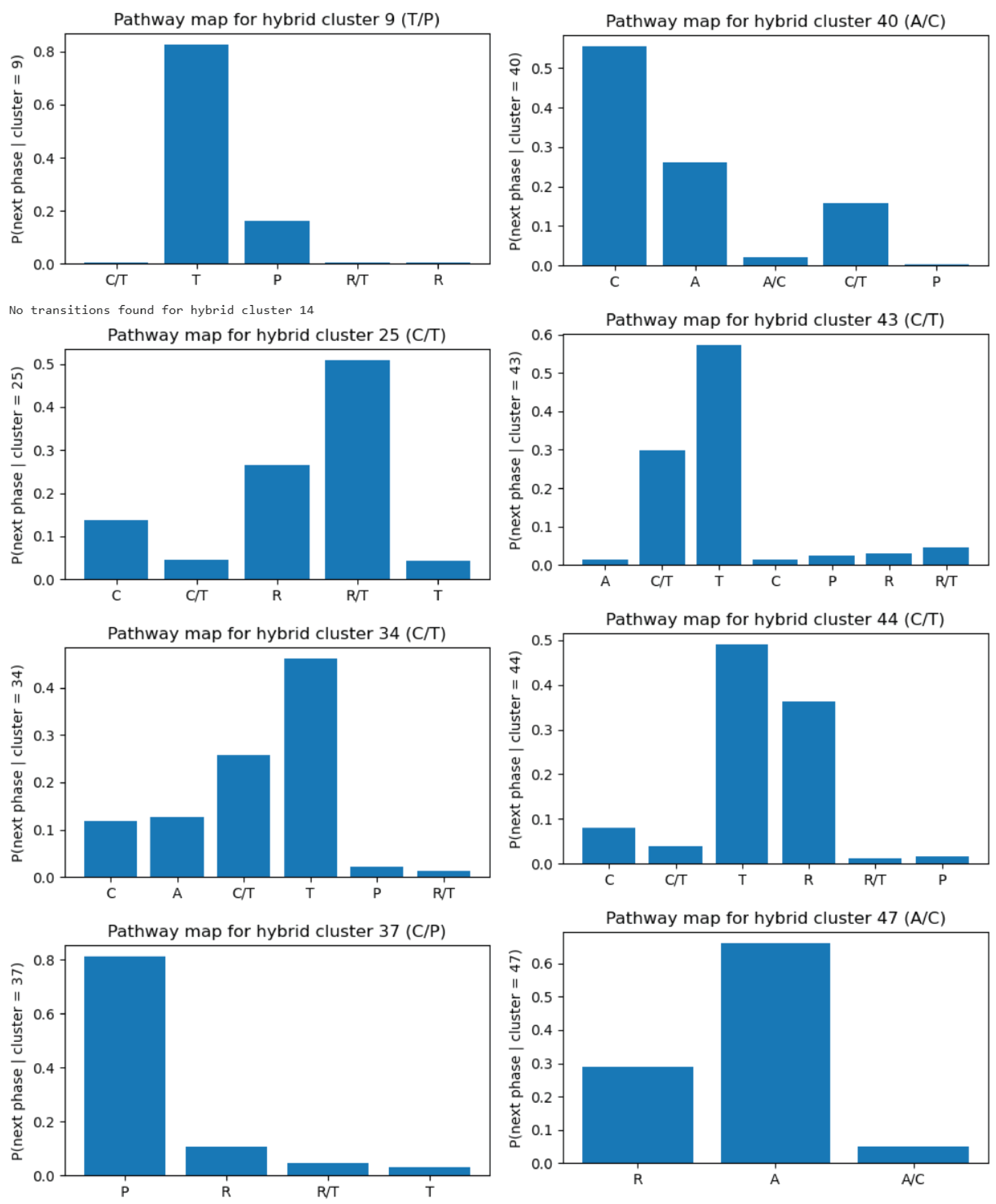}
\caption{\label{fig:figs3} Transition pathways for each hybrid cluster, showing the distribution of next‑phase outcomes. These panels summarize the dominant routes of hybrid states and complement the full transition analysis presented in Fig. 6.}
\end{figure*}

\begin{figure*}[!ht]
\centering
\includegraphics[width=0.9\textwidth]{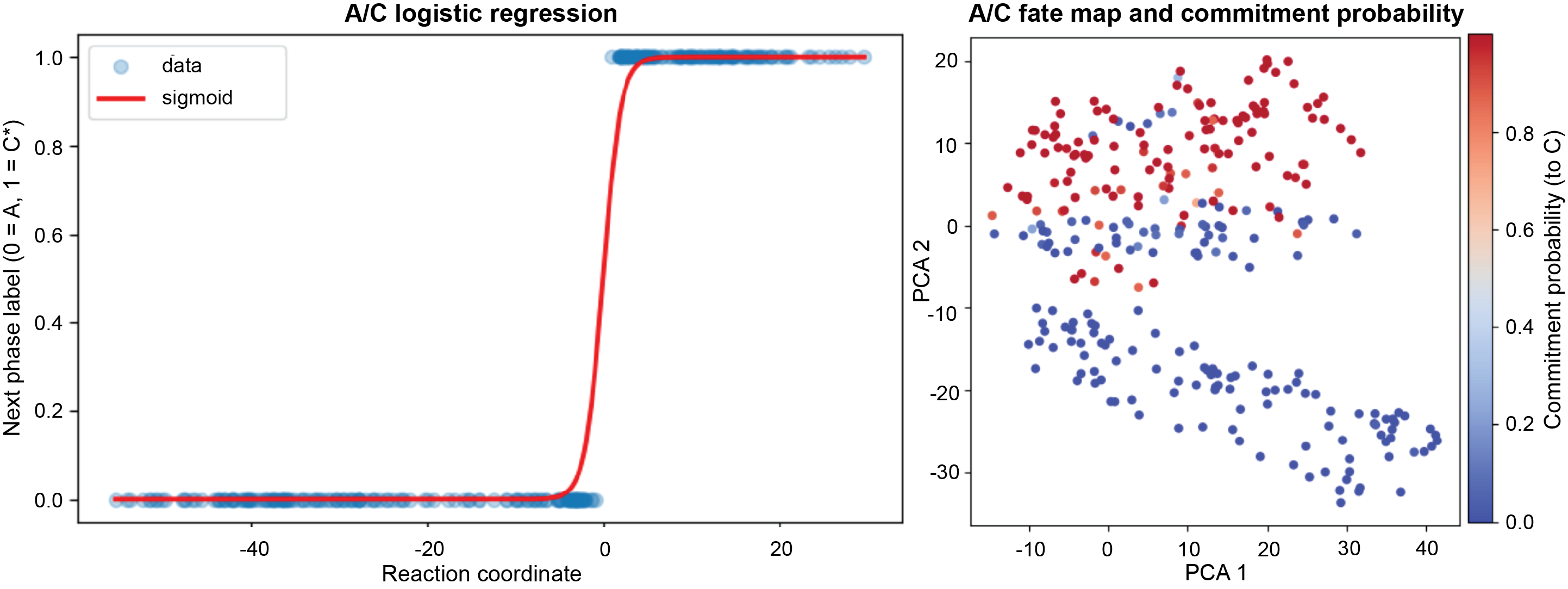}
\caption{\label{fig:figs4}  Logistic regression analysis of A/C hybrid fate. The reaction coordinate separates dissolving (A) from crystallizing (C) A/C states, and the PCA fate map colored by the logistic-regression commitment probability $q$ reveals a latent space gradient in which commitment to C is sharply localized.}
\end{figure*}

\begin{table*}[!h]
\centering
\renewcommand{\arraystretch}{1.0}
\setlength{\tabcolsep}{10pt}

\begin{tabular}{l|p{6cm}}
\hline
\textbf{Phase Class} & \textbf{Cluster Indices} \\
\hline

A (Amorphous) 
& 0, 3, 6, 12, 18, 28, 29, 39, 48 \\

C (Crystalline) 
& 1, 5, 10, 11, 16, 17, 27, 35, 38, 46, 49 \\

R (Recrystallized) 
& 2, 4, 7, 13, 20, 26, 32 \\

P (Polycrystalline) 
& 19, 23, 24, 30, 31, 36, 41, 45 \\

T (Twinned) 
& 8, 15, 21, 22, 33, 42 \\

Hybrid States 
& 
A/C: 40, 47; \\
& C/T: 25, 34, 43, 44; \\
& C/P: 37; \\
& T/P: 9; \\
& R/T: 14 \\
\hline
\end{tabular}

\caption{Cluster-to-phase classification for all 50 latent-space clusters. Clusters are grouped into structural classes based on diffraction signatures and latent-space geometry. Hybrid states represent intermediate structural motifs lying between major basins.}
\label{tab:cluster_phase_map}
\end{table*}

\begin{table*}[!h]
\centering
\renewcommand{\arraystretch}{1.2}
\setlength{\tabcolsep}{12pt}

\begin{tabular}{c|ccccc}
\hline
 & \multicolumn{5}{c}{\textbf{Full Temperature Range}} \\
\textbf{From/To} 
& \textbf{A} & \textbf{C$^\ast$} & \textbf{H} & \textbf{P} & \textbf{T} \\
\hline

A      
& 0.77 & 0.08 & 0.08 & 0.02 & 0.04 \\

C$^\ast$ 
& 0.03 & 0.89 & 0.07 & 0.00 & 0.01 \\

H      
& 0.10 & 0.21 & 0.20 & 0.12 & 0.37 \\

P      
& 0.00 & 0.00 & 0.00 & 0.94 & 0.06 \\

T      
& 0.00 & 0.01 & 0.08 & 0.05 & 0.86 \\

\hline
\end{tabular}

\caption{Phase transition probabilities combined across the full temperature range. Crystalline (C) and recrystallized (R) states are combined into a single ordered basin C$^\ast$. The matrix reveals a strongly absorbing P and T basin at high temperature, a stable amorphous basin at low temperature, and hybrid-mediated pathways that connect these regimes.}
\label{tab:msm_full}
\end{table*}

\begin{table*}[t]
\centering
\footnotesize
\setlength{\tabcolsep}{3pt}
\renewcommand{\arraystretch}{1.1}

\begin{minipage}[t]{0.48\textwidth}
\centering
\textbf{Temp Steps 0--1}

\begin{tabular}{c|ccccc}
\hline
From/To & A & C$^\ast$ & H & P & T\\
\hline
A       & 0.98 & 0.01 & 0.01 & 0.00 & 0.00\\
C$^\ast$& 0.07 & 0.85 & 0.07 & 0.00 & 0.00\\
H       & 0.34 & 0.56 & 0.10 & 0.00 & 0.00\\
P       & 0.00 & 0.00 & 0.00 & 0.00 & 0.00\\
T       & 0.00 & 0.00 & 0.00 & 0.00 & 0.00\\
\hline
\end{tabular}

\vspace{1em}

\textbf{Temp Steps 2--3}

\begin{tabular}{c|ccccc}
\hline
From/To & A & C$^\ast$ & H & P & T \\
\hline
A       & 0.91 & 0.09 & 0.00 & 0.00 & 0.00 \\
C$^\ast$& 0.14 & 0.82 & 0.04 & 0.00 & 0.00 \\
H       & 0.89 & 0.06 & 0.06 & 0.00 & 0.00 \\
P       & 0.00 & 0.00 & 0.00 & 0.00 & 0.00 \\
T       & 0.00 & 0.00 & 0.00 & 0.00 & 0.00 \\
\hline
\end{tabular}

\vspace{1em}

\textbf{Temp Steps 4--5}

\begin{tabular}{c|ccccc}
\hline
From/To & A & C$^\ast$ & H & P & T \\
\hline
A       & 0.89 & 0.06 & 0.05 & 0.00 & 0.00 \\
C$^\ast$& 0.03 & 0.80 & 0.16 & 0.00 & 0.00 \\
H       & 0.75 & 0.00 & 0.25 & 0.00 & 0.00 \\
P       & 0.00 & 0.00 & 0.00 & 0.00 & 0.00 \\
T       & 0.00 & 0.00 & 0.00 & 0.00 & 0.00 \\
\hline
\end{tabular}

\vspace{1em}

\textbf{Temp Steps 6--7}

\begin{tabular}{c|ccccc}
\hline
From/To & A & C$^\ast$ & H & P & T \\
\hline
A       & 0.00 & 0.00 & 0.38 & 0.62 & 0.00 \\
C$^\ast$& 0.00 & 0.85 & 0.02 & 0.05 & 0.01 \\
H       & 0.01 & 0.04 & 0.25 & 0.03 & 0.67 \\
P       & 0.00 & 0.78 & 0.07 & 0.02 & 0.13 \\
T       & 0.00 & 0.01 & 0.05 & 0.02 & 0.92 \\
\hline
\end{tabular}

\end{minipage}
\hfill
\begin{minipage}[t]{0.48\textwidth}
\centering

\textbf{Temp Steps 1--2}

\begin{tabular}{c|ccccc}
\hline
From/To & A & C$^\ast$ & H & P & T \\
\hline
A       & 0.80 & 0.20 & 0.00 & 0.00 & 0.00 \\
C$^\ast$& 0.49 & 0.39 & 0.12 & 0.00 & 0.00 \\
H       & 1.00 & 0.00 & 0.00 & 0.00 & 0.00 \\
P       & 0.00 & 0.00 & 0.00 & 0.00 & 0.00 \\
T       & 0.00 & 0.00 & 0.00 & 0.00 & 0.00 \\
\hline
\end{tabular}

\vspace{1em}

\textbf{Temp Steps 3--4}

\begin{tabular}{c|ccccc}
\hline
From/To & A & C$^\ast$ & H & P & T \\
\hline
A       & 0.95 & 0.03 & 0.02 & 0.00 & 0.00 \\
C$^\ast$& 0.07 & 0.93 & 0.00 & 0.00 & 0.00 \\
H       & 0.00 & 0.98 & 0.02 & 0.00 & 0.00 \\
P       & 0.00 & 0.00 & 0.00 & 0.00 & 0.00 \\
T       & 0.00 & 0.00 & 0.00 & 0.00 & 0.00 \\
\hline
\end{tabular}

\vspace{1em}

\textbf{Temp Steps 5--6}

\begin{tabular}{c|ccccc}
\hline
From/To & A & C$^\ast$ & H & P & T \\
\hline
A       & 0.00 & 0.20 & 0.31 & 0.21 & 0.27 \\
C$^\ast$& 0.00 & 0.92 & 0.05 & 0.00 & 0.03 \\
H       & 0.00 & 0.75 & 0.23 & 0.00 & 0.02 \\
P       & 0.00 & 1.00 & 0.00 & 0.00 & 0.00 \\
T       & 0.00 & 0.50 & 0.25 & 0.00 & 0.25 \\
\hline
\end{tabular}

\end{minipage}
\caption{Collapsed transition matrices for all consecutive temperature steps.}

\end{table*}
